\documentclass[a4paper,12pt]{article}
\usepackage{amsfonts,amsthm,amsmath,amssymb,graphicx,relsize}

\newcommand{\cA}{\mathcal A}

\newcommand{\cF}{\mathcal F}

\newcommand{\cM}{\mathcal M}

\newcommand{\cP}{\mathcal P}

\newcommand{\be}{\begin{equation}}
\newcommand{\bea}{\begin{eqnarray}}
\newcommand{\ee}{\end{equation}}
\newcommand{\eea}{\end{eqnarray}}

\newcommand{\ccL}{ \mathcal{L}} 

\def\mC{ \mathbb{C}}

\def\Li{ \hbox{Li} } 

\DeclareMathOperator*{\LLim}{Lim}

\begin{document} 

\rightline{QMUL-PH-15-23}

\vspace{1.8truecm}

{\LARGE{ 
\centerline{\bf  Interactions as conformal  intertwiners in 4D QFT } 
}}  

\vskip.5cm 

\thispagestyle{empty} \centerline{
    {\large \bf Robert de Mello Koch
${}^{a,} $\footnote{ {\tt robert@neo.phys.wits.ac.za}}}
   {\large \bf and Sanjaye Ramgoolam
               ${}^{b,}$\footnote{ {\tt s.ramgoolam@qmul.ac.uk}}   }
                                                       }

\vspace{.4cm}
\centerline{{\it ${}^a$ National Institute for Theoretical Physics ,}}
\centerline{{\it School of Physics and Mandelstam Institute for Theoretical Physics }}
\centerline{{\it University of Witwatersrand, Wits, 2050, } }
\centerline{{\it South Africa } }

\vspace{.4cm}
\centerline{{\it ${}^b$ Centre for Research in String Theory, School of Physics and Astronomy},}
\centerline{{ \it Queen Mary University of London},} \centerline{{\it
    Mile End Road, London E1 4NS, UK}}

\vspace{1truecm}

\thispagestyle{empty}

\centerline{\bf ABSTRACT}

\vskip.2cm

In a recent paper we showed that the correlators of free scalar field theory in four dimensions 
can be constructed from a two dimensional topological field theory based on $so(4,2)$ equivariant maps (intertwiners). The free field result, along with   results of Frenkel and Libine on equivariance properties of Feynman integrals,  are developed further in this paper. We show that the coefficient of the log term in the 1-loop   4-point conformal integral is a projector in the tensor product of  
$so(4,2)$ representations. We also show that the 1-loop 4-point integral can be written as a sum of four terms, each associated with the quantum equation of motion for one of  the four external legs. The quantum equation of motion is shown to be related to equivariant maps involving indecomposable representations of $ so(4,2)$, a phenomenon which illuminates  multiplet recombination. The harmonic expansion method for Feynman integrals  is a powerful tool for arriving at these results. The generalization to other interactions and  higher loops is discussed.

\setcounter{page}{0}
\setcounter{tocdepth}{2}

\newpage

\tableofcontents

\setcounter{footnote}{0}

\linespread{1.1}
\parskip 4pt

{}~
{}~

\section{Introduction}
\label{sec:intro}

Many aspects of the combinatorics of $ N=4$ super-Yang-Mills theories
 have been shown to be captured by two-dimensional topological field theories (TFT2s)  based on permutation groups \cite{Feyncount,DMW, QuivCalc,KimFrob,Tom-complex,CJR,GiGraOsc,DoubCos,Geloun:2013kta}. 
Specifically these topological 
field theories were found in the  computation of    correlators in the free limit of gauge 
theories, the enumeration of states for open strings connecting branes and the 
construction of their wavefunctions diagonalizing  the 1-loop dilatation operator \cite{minzar,bks0303},  the enumeration of Feynman diagrams
and tensor model observables. In the context of $ N=4 $ SYM correlators, 
this leads naturally to the question of whether 
the space-time dependences of correlators (as well as the combinatoric dependences
on the operator insertions) can be captured by an appropriate TFT2. As a simple 
test case to explore this question, we showed that free scalar field correlators 
in four dimensions can be reproduced by a TFT2 with $ so(4,2)$ invariance \cite{CFT4TFT2}. We used Atiyah's  axiomatic framework for TFT2s, where tensor products of a state space are associated with disjoint unions of circles and linear 
homomorphisms are associated with interpolating surfaces (cobordisms) \cite{Atiyah}.
The properties of cobordisms in two dimensions are reflected in the algebraic structure of a Frobenius algebra : an associative algebra with a non-degenerate pairing. The notion of a TFT2 with global $G$ symmetry was given in \cite{mooseg} : The state spaces are representations of the group $G$ and 
the linear maps are equivariant with  respect to the $G$-action.

In the construction of \cite{CFT4TFT2}, the basic two-point function in scalar field theory is related to 
the invariant map $ (  V  \otimes V   ) \rightarrow \mC $, where $V$ 
is a direct sum of two irreducible representations of $so(4,2)$. 
The irreducible representation (irrep)  $V_+$ contains a lowest weight state 
corresponding to the basic scalar field via the operator-state correspondence 
\bea 
v^+ = \LLim_{ |x|  \rightarrow 0 } \phi ( x ) | 0 > 
\eea 
Translation operators ( $P_{\mu} $ )  act on the lowest weight state to 
generate a tower of states. They can be viewed as raising operators since 
\bea 
[ D , P_{ \mu} ] = P_{ \mu} 
\eea
The state  $ P_{\mu} P_{\mu} v^+  $ is set to zero to correspond to the equation of motion
of the scalar. The general state in this representation is 
\bea 
Y^{l}_{m} ( P ) v^+ 
\eea 
where $Y^{ l }_{ m }(P)$ is a symmetric traceless tensor of $so(4)$ 
contracted with a product of $P$'s. The integer $l$ gives the degree of  the  polynomial in $P$ and $m$ labels a state in the symmetric traceless tensor representation.  We will refer to $V_+$ as a positive energy representation, a terminology inspired 
by AdS/CFT where the scaling dimension in CFT is energy for global time translations in AdS 
\cite{malda,gkp,witten}. 
The irrep $ V_- $ is dual to $V_+$. It contains a dual state $v^-$ of dimension $-1$ 
and other states are generated by acting with symmetric traceless combinations of $ K_{ \mu} $
\bea
Y^{l}_m ( K ) v^-
\eea
There is a non-degenerate invariant pairing $ \eta  : V_+ \otimes V_- \rightarrow \mC $.
We refer to $ V_-$ as a negative energy representation, since it contains states with negative dimension. 

The foundation of the  TFT2 approach to free CFT4 correlators is to consider 
the local  quantum field at $x$ as a state in $V_+ \oplus V_-$ 
\bea\label{thequantumfield} 
\Phi ( x ) = { 1 \over \sqrt { 2 } } \left (  e^{ - i P \cdot x } v^+ + (x')^2 e^{ i K\cdot x' } v^- \right )  \equiv \Phi^{+} + \Phi^- 
\eea
Here $ x'^{ \mu} = { x^{\mu} \over x^2} $. It is found that 
\bea 
\eta ( \Phi ( x_1  ) , \Phi ( x_2 ) ) = { 1 \over ( x_1 - x_2 )^2 } 
\eea
We can think of $ e^{ -  i P \cdot x } $ and $ e^{ i K\cdot x' } $ as four dimensional 
analogs of the two dimensional vertex operators familiar from string theory and 2D conformal field theory. 
In 2D CFT physical states of the string are constructed from exponentials of the coordinate  quantum fields $X^{\mu}$
which have an expansion in oscillators coming from quantizing string motions. In the case of 
CFT4/TFT2 at hand,  the exponential is in the momentum operators (and the special conformal translations which are related to the momenta by inversion), which are among the generators of $so(4,2)$.   Other developments in CFT4 inspired by vertex operators include
\cite{KKN1410,JKPS1410,BKV1505,ADGN0510}.  It is intriguing that the 2D CFT vertex operators have the   coordinate as an operator in the exponential, whereas here we are using the momenta as operators in the exponential.  Conceivably there is some form of $x-p$ duality between these different types of vertex operators. Clarifying this  could be useful in understanding the role of Born reciprocity (a theme revived  recently in \cite{FLM1405,FLM1502})  in strings and QFT. It is also worth pointing out other earlier developments on 
higher dimensional generalizations of 2D vertex operators \cite{banks,moore1,moore2}.

The realization of CFT4 correlators in terms of TFT2 means that we are writing quantum field theory correlators in 
terms of standard representation theory constructions. CFT4/TFT2 builds on but 
goes beyond the standard use of   representation theory as a tool 
to calculate quantities defined by a path integral. Rather it is a reformulation of correlators of a quantum field theory 
in terms of standard constructions of representation theory, notably linear representations and equivariant maps between them. 
The appearance of both positive and negative energy representations in (\ref{thequantumfield}) is an important part of this 
reformulation. For example, while the free field OPE 
\bea 
\phi^2 ~\times  ~   \phi^2 \rightarrow \phi^4
\eea
could be understood by using an expression for $\phi$ in terms of strictly positive energy representations, this is not  the case for 
\bea 
\phi^2 ~ \times ~  \phi^2 \rightarrow \phi^2 
\eea
The latter involves the invariant map $ \eta $ contracting a positive and a negative energy representation. 
This linearizes the CFT4 by relating correlators to linear equivariant maps.
The construction achieves this by passing from the space of operators built on the primary at $x=0$ to 
 the ``doubled space'' of operators, built on the primary at $x=0$ (positive energy) and $x=\infty$ (negative energy).

This paper addresses the natural question of whether the  free field construction of 
\cite{CFT4TFT2} is relevant to perturbative quantum field theory.  We explain this question in more technical terms 
in section \ref{sec:backgd} and show how it leads to the expectation that conformal integrals should be related to 
intertwiners involving representations of $so(4,2)$.  These integrals are important building blocks in perturbation theory 
\cite{Dixon-review-1310} and have been shown recently to have remarkable properties, called magic identities \cite{DHSS07,DKS0707}.
Interestingly, equivariance properties of the kind suggested by CFT4/TFT2 have already  been found 
in work of Frenkel and Libine\cite{FL07}, who were approaching Feynman integrals from the point of view of quaternionic analysis.
Group-theoretic interpretations of relativistic holography have also been suggested, through the explicit 
construction of the boundary-to-bulk operators for arbitrary integer spin as intertwining operators\cite{Aizawa:2014yqa}.
The physics literature on higher dimensional conformal blocks suggests equivariance properties of these integrals, notably 
the works of \cite{DO0006,DO0011,SD1204} which approach the conformal blocks in terms of Casimir differential 
equations and subsequent reformulation in terms of the shadow formalism. As indicated  by  the discussion of OPEs above, 
 the QFT discussions of conformal blocks 
do not immediately imply an interpretation in terms of linear representations and associated equivariant maps. 
However, the use of Casimir differential equations is a powerful tool in arriving at the equivariant 
map interpretation of QFT quantities. 
The exponential vertex operators play an important role in what follows because they allow us to map algebraic generators 
of the $ so(4,2)$ Lie algebra to differential operators acting on function spaces.  
In particular, the Casimirs in the (universal enveloping)
$so(4,2)$ algebra become Casimir differential operators. 

Section \ref{sec:backgd} reviews some aspects of the work of Frenkel-Libine which  we will find useful 
in developing the vertex operator approach to 
these equivariant maps. We also review here some basic facts about indecomposable 
representations which will be useful for section \ref{sec:QEOM-indec}. 
In this paper, our primary focus is on the conformal 4-point integral, whose exact answer is known  \cite{UssDavPLB305,UssDav9307}. 
Our first main result is that the coefficient of the log-term in the 4-point answer is given by 
the matrix elements of an equivariant map $ V_+ \otimes V_+ \otimes V_- \otimes V_- \rightarrow \mC $. 
Section \ref{sec:HPEM} reviews the harmonic expansion method which is used to arrive at this result. 
This method involves the expansion of the propagator in terms of $so(4)$ harmonics. For a given order of 
the external points in the conformal integral ($ |x_1| < |x_2| < | x_3| < |x_4| $), we separate the integral into 
regions according to the range of integration of $|x|$. 
One region  $ |x_2| < |x| < |x_3|$ leads to the logarithmic term. The result that the coefficient of the logarithmic term is 
an intertwiner is derived in Section \ref{sec:CoeffLog}. This section contains our first main result, equation (\ref{theclaim}). 
 Appendix \ref{SumClebschs} explains how the above 
result leads to an identity for an infinite sum of products of $su(2)$ Clebsch-Gordan coefficients. 

In section \ref{sec:QEOM-indec} we will consider the other regions of integration and show they 
can be collected into four different terms associated with  the quantum equation of motion 
for each of the external variables $x_i$. On each of the terms, the action of Laplacians  gives $so(4,2)$ invariant equivariant maps involving a submodule of  these indecomposable representations. 
For two of the four terms, the equivariant maps employ the standard co-product and we show how they can be lifted
from the sub-module to the full indecomposable representation. The remaining two terms make use of a twisted co-product. In these cases, we believe  the lift to the full indecomposable representation is possible but there are technical subtleties which remain to be clarified.
These results show  that the full integral can be viewed 
as an equivariant map obtained by lifting from the sub-module to the full indecomposable representation. Equation (\ref{main2}) is the second main result of this paper.  It links a beautiful structure in representation theory to quantum equations of motion arising from  the collision of 
interaction point with external points, the source of many deep aspects of quantum field theory.
 The appearance of indecomposable representations is closely related to multiplet recombination. This phenomenon, in connection 
with quantum equations of motion and the Wilson-Fischer fixed point,  has also recently 
been discussed \cite{Rych-Tan}. Recombination of superconformal multiplets has also been 
extensively discussed in the context of  $N=4$ and $N=2$ theories (see for example \cite{DO0209,KMMR0512,bouffe}
and refs therein), the breaking of higher spin symmetry in AdS/CFT being one of the motivations. 

In the final section, we outline how our results extend to higher loops and describe  other 
future directions of research.

\section{ Background and motivations   } 
\label{sec:backgd} 

\subsection{ CFT4/TFT2 suggests equivariant interpretation of perturbative  Feynman integrals } 

Once we have a formulation of all the correlators in free CFT4 
in terms of TFT2 of equivariant maps, the natural question is : Can we 
describe perturbation theory away from free CFT4 in the language of the 
TFT2 ? Since perturbation theory involves the integration of  correlators in the free field 
 theory, weighted with appropriate powers of coupling constants, once we have 
a TFT2 description of all the free field correlators, we are part of the way there. 
The important new ingredient is integration of the interaction vertices whose 
consistency with  equivariant maps remains to be established. A natural place to 
start this investigation is the  case of conformal integrals \cite{DHSS07,DKS0707} involving scalar fields.
It is known that general perturbative integrals in four dimensions at one-loop can be reduced to a basis of 
scalar integrals, involving the box, the triangle and bubble diagrams (see \cite{Dixon-review-1310} and refs therein).  
The momentum space box diagram becomes, after Fourier transformation to coordinate space, a 
diagram related by graph duality to the original graph. The integral of interest in coordinate space is 
\bea\label{IntegralOfInterest} 
I ( x_1 , x_2 , x_3 , x_4 ) = \int {d^4 x\over 2\pi^2} { 1 \over  ( x_1 - x)^2  ( x_2 - x)^2 ( x_3 - x)^2 ( x_4 - x )^2 } 
\eea 
This integral (\ref{IntegralOfInterest}), viewed as the kernel of an integral operator, acting on appropriate test functions, 
has been shown to be related to equivariant maps in \cite{FL07}.  There are two distinct equivariant 
interpretations developed there : one involves the Minkowski space integral, and the other 
involves integration over a $U(2)$ in complexified space-time. Subsequent higher loop 
generalizations have been given \cite{Lib1309,Lib1407}.

Here we give a qualitative explanation of how  the TFT2 way of thinking about perturbation theory suggests an 
equivariant interpretation for integrals.
Subsequently we will investigate the expectations directly.

We can choose all the external vertex operators to be 
\bea\label{extlegs1}  
(x_1')^2 e^{ iK\cdot x_1' } v^- \otimes  (x_2')^2 e^{ iK\cdot x_2' } v^- \otimes (x_3')^2 e^{ iK\cdot x_3' } v^-
 \otimes (x_4')^2 e^{ iK\cdot x_4' } v^-
\eea
Take a tensor product with 
\bea\label{intlegs1}  
e^{ - iP\cdot x } v^+ \otimes  e^{ - iP\cdot x } v^+ \otimes  e^{ - iP\cdot x } v^+ \otimes  e^{ - iP\cdot x } v^+
\eea
Take a product of $ \eta $ pairings between the first factor in (\ref{extlegs1}) with  the first 
factor in (\ref{intlegs1}), the second with second etc.   This produces the product of propagators in 
(\ref{IntegralOfInterest}). 
In another way to set up the correlator, use as external states 
\bea\label{MiddleEquiv}  
e^{ - iP\cdot x_1 } v^+ \otimes  e^{ - iP\cdot x_2 } v^+ \otimes x_3'^2 e^{ iK\cdot x_3' } v^- 
\otimes  x_4'^2 e^{ iK\cdot x_4' } v^- 
\eea
To this we tensor 
\bea\label{MiddleEquiv2}  
(x')^2 e^{ - iK\cdot x' } v^+ \otimes  (x')^2  e^{ - iK\cdot x' } v^+ \otimes  e^{ - iP\cdot x } v^+ \otimes  e^{ - iP\cdot x } v^+
\eea
Again we pair  the $i$'th factor in (\ref{MiddleEquiv}) with the corresponding factor in (\ref{MiddleEquiv2}). 
All the internal vertex operators have a common space-time position, which is integrated over. The integrands can be reproduced by the TFT2 method.

The different choices for external vertex operators should correspond to expansions in positive 
powers of $x_i$ or of $x_i' = { x_i \over |x_i^2|} $.  A method of integration which connects 
with the above vertex operator method of thinking about the integral is known 
as the Harmonic Polynomial Expansion Method (HPEM), which give formulae that can be simplified 
using Gegenbauer polynomials \cite{Kotikov}. We will choose an ordering of the external points 
$ |x_1| < |x_2| < |x_3| < |x_4| $ and do the integral in Euclidean space, separating 
it into five parts depending on the range of $|x|$. For each range we will apply the HPEM. 

The choice (\ref{MiddleEquiv}) corresponds to the region $   |x_2| < |x| < |x_3|  $, where we 
will find a logarithmic term. There are no logs from any of the other regions. This follows from 
basic group theoretic properties of $so(4) = su(2) \times su(2)$ tensor products, when these 
are used in conjunction with the HPEM. We will describe this in more detail in Section \ref{sec:HPEM}.
For now we notice that the natural quantity to look at in search of an equivariant 
interpretation is $ x_3^2 x_4^2 I ( x_1 , x_2 , x_3 , x_4 ) $. In section \ref{sec:CoeffLog} 
we will establish that the coefficient of the log term in $ x_3^2 x_4^2 I ( x_1 , x_2 , x_3 , x_4 ) $
can indeed be interpreted in terms of an equivariant map. In arriving at this we will make contact 
with the results of \cite{FL07}, in particular their discussion of a version of the integral where the 
contour of integration is taken to be a copy of $ U(2)$ instead of Minkowski space. 

In section \ref{sec:QEOM-indec} we will consider the other regions of integration and show they 
can be collected into four different terms associated with  the quantum equation of motion 
for each of the external variables $x_i$. This separation will be used to give 
an interpretation in terms of equivariant maps for the full integral. 

\subsection{Conformal integral : Exact answer and an expansion }\label{CIandFL} 

The integral (\ref{IntegralOfInterest}) belongs to a class of conformal integrals which have been exactly solved. 
In momentum space, the integral is a 1-loop box. 
The result is \cite{UssDavPLB305,UssDav9307}
\bea 
I ( x_1 , x_2 , x_3 , x_4 )  = {1\over 2 x_{ 13}^2 x_{24}^2} \Phi ( s,  t )\label{exactresult} 
\eea
where
\bea
\Phi ( s , t ) = { 1 \over \lambda } \left (  2  ( \Li_2 ( - \rho s ) + \Li_2 ( - \rho t ) ) +  \ln ( \rho s ) \ln ( \rho t ) 
+ \ln \left({ t \over s }\right) \ln \left({ 1 + \rho t   \over 1 + \rho s  }\right)
                         + { \pi^2 \over 3 }                                 \right ) 
\eea
and 
\bea 
&& \rho =  { 2 \over 1 - s - t + \lambda } \hspace*{1cm} \lambda =   \sqrt { ( 1 - s - t)^2 - 4 st } \cr 
&& s = { x_{12}^2 x_{34}^2 \over x_{13}^2 x_{24}^2 } \hspace*{1cm} 
     t = { x_{14}^2 x_{23}^2 \over   x_{13}^2 x_{24}^2 } 
\eea
We will need the expansion of $\Phi$ about $s=0$ and $t=1$.
Towards this end we introduce $t=1+u$ and take the limit $s \rightarrow 0$ first and then $u \rightarrow 0$.
In this limit 
\bea 
&&  \lambda \sim u + s ( -1 - { 2 \over u } ) + s^2 ( - { 2 \over u^3 } - { 2 \over u^2 } ) \sim u + s ( -1 - { 2 \over u } )   \cr 
&&  \rho \sim s^2 \left(\frac{2}{u^5}+\frac{1}{u^4}\right)
+\frac{s}{u^3}+\frac{u^2-u}{s}-u^2+u+\frac{1}{u}-1 \sim \frac{u^2-u}{s}
\eea
In the limit we consider, since $\rho \rightarrow \infty$, we need to apply the identity
\bea
    \text{Li}_2 (z)=-\text{Li}_2(1/z)-{\pi^2\over 6}-{1\over 2} \log^2 (-z)
\eea
to rewrite $\text{Li}_2 ( \rho t )$. 
%
%
After this transformation
\bea 
\Phi( s , t ) = { 1 \over \lambda } \left \{ 2 \Li_2 ( -\rho s ) -  2 \Li_2 ( - \rho^{-1}  t^{-1}  ) - \log \left (  {s \over t } \right )   \log \left ( {    ( 1 + \rho^{-1} t^{-1} ) \over ( 1 + \rho s ) }               \right )  \right \} 
\eea
Since we will discuss the coefficient of the log extensively in what follows, we introduce the notation
\begin{align*} 
& \Phi ( s , u  )  = F_0 ( s , u ) + \log (s) F_1 ( s  , u ) \cr 
& F_0 ( s , u ) = { 1 \over \lambda } \left \{  2 \Li_2 ( - \rho s ) - 2 \Li_2 ( - \rho^{-1} ( 1+u)^{-1} ) + \log (  1 + u )    
\log \left( { (1+\rho^{-1} t^{-1} ) \over ( 1 + \rho s ) } \right )  \right  \} \cr 
& F_1 ( s , u ) = - {1\over\lambda}\log \left ( {( 1 + \rho^{-1} t^{-1} ) \over (1+\rho s )} \right ) 
\end{align*} 
We are interested in the limit $ |x_2| > |x_1| \rightarrow 0 $ with $ |x_4| > |x_3| \rightarrow\infty$. 
This means that $x_1 , x_2 \rightarrow 0,$  $x_3' , x_4' \rightarrow 0 $.  In this limit 
\begin{align*}  
& s = { ( x_1 - x_2 )^2 ( x_3 - x_4)^2 \over ( x_1 - x_3)^2 ( x_2 - x_4)^2 }  \cr 
& ={  ( x_1 - x_2 )^2 ( x_3' - x_4')^2 \over f ( x_1 , x_3') f ( x_2 , x_4')  } \cr 
& t = { ( x_1 - x_4)^2 ( x_2 - x_3)^2 \over ( x_1 - x_3 )^2 ( x_2 - x_4)^2 } \cr 
& = { f ( x_1 , x_4' ) f ( x_2 , x_3' ) \over f ( x_1 , x_3') f ( x_2 , x_4') } 
\end{align*}
where
\bea 
f ( x , y' )  =1 + 2 x\cdot y' + x^2 y'^2 
\eea
These equations show that $u$ and $s$ are real-analytic in the limit, admitting expansions in 
$ x_1 , x_2 , x_{3}' , x_4'$. While  $ \lambda , \rho $ do not have an expansion in positive powers of 
$ s , u $ as $ s, u \rightarrow 0$, the quantity $ F_1 (s , u ) $ does have such an expansion. This leads to an expansion of $ F_1 ( s ( x_1, x_2 , x_3' , x_4')  , u ( x_1 , x_2 , x_3' , x_4')  ) $ in powers  of $ x_1 , x_2 , x_3' , x_4' $
will be related to a projector in section \ref{sec:CoeffLog}.  

\subsection{Indecomposable representations and multiplet recombination  } 

We will review the notion of indecomposable representations and explain their 
relevance to the recombination of multiplets when interactions are turned on. 

As a simple example, consider the Lie algebra $su(2)$ with generators $ J_3 , J_{ \pm}$. 
\bea 
&& [ J_{3} , J_{\pm} ] = \pm J_{ \pm } \cr 
&& [ J_+ , J_- ] = 2 J_3
\eea
With this normalization of the generators, irreducible representations have
$J_3$ eigenvalues in the range $ \{ j ,  j-1 , \cdots , - j \} $ for $ j \in \{ 0 , { 1 \over 2 } , 1 , { 3 \over 2 } , \cdots \} $. 
Consider  a lowest weight representation built by starting with  a state $ | - { 1 \over 2 } \rangle $ satisfying 
\bea 
J_3 |-  { 1 \over 2 } \rangle  &=&  -  { 1 \over 2 } | - { 1\over 2 } \rangle \cr 
J_- | - {  1 \over 2 } \rangle &=&  0 
\eea
Now consider the infinite dimensional representation spanned by $ J_+^n| - { 1\over 2 } \rangle  $ 
for $ n \in \{ 0 , 1, 2 , \cdots \} $. Denote this representation by $ \widetilde V_{-  {1\over 2} } $. 
The state $ J_+^2  | -  {  1 \over 2 } \rangle $ has the property that it is annihilated by $ J_-$
\bea 
J_- J_+^2   | { -  1 \over 2 } \rangle  =  0
\eea
This has the consequence that the vector subspace of $ \widetilde V_{ - { 1 \over 2 } } $ 
spanned by $ J_+^n | - { 1/2} > $ for $n \ge 2 $ is an invariant subspace of $ \widetilde V_{ - { 1 \over 2} } $. 
Denote this subspace as $ V_{- { 1 \over 2} }^{ (2)} $. 
The quotient space $ \widetilde V_{ - { 1 \over 2 } }/  V_{- { 1 \over 2 } }^{ (2)}  $ 
is the standard  two-dimensional representation of $ su(2)$. We have an exact sequence 
\bea 
0 \rightarrow V_{  - { 1 \over  2 } }^{(2)}  \rightarrow \widetilde V_{ - { 1 \over 2 } }   
\rightarrow \widetilde V_{-{ 1 \over 2 } } / V_{  - {  1 \over  2 } }^{(2)}  \rightarrow 0   
\eea
The quotient space admits a positive definite inner product. If we choose an inner product where $ | -{ 1\over 2 } \rangle$ 
has unit norm, then  $ J_+^2 |- { 1 \over 2 } \rangle $ has zero norm. Setting this null state to zero gives the quotient space 
which is a unitary representation of $su(2)$. 

In four dimensional free scalar quantum field theory, we encounter the representation $ V_+$ containing  a lowest weight state  
$v^+$ of dimension $1$. There are additional states of higher dimension of the form 
\bea 
T_I^{ \mu_1 \mu_2 \cdots \mu_n } P_{ \mu_1} \cdots P_{ \mu_n } v^+ 
\eea
where the $T_I$ are symmetric traceless tensors. This is a unitary representation of $so(4,2)$. 
By direct analogy to the above discussion, $V_+$ is obtained as a quotient space of a larger representation 
$ \widetilde V_+$ spanned by
\bea 
S_{I}^{ \mu_1 \cdots \mu_n }  P_{ \mu_1} \cdots P_{ \mu_n } v^+ 
\eea
where the $ S_I$ are symmetric tensors (not necessarily traceless). To get to $ V_+$, we quotient $ \widetilde V_+ $ 
by the subspace spanned by 
\bea 
S_{I}^{ \mu_1 \cdots \mu_n }  P_{ \mu_1} \cdots P_{ \mu_n }   P_{ \mu } P_{ \mu } v^+ 
\eea
Denoting this subspace by $ V_+^{ (p^2)} $ we have the exact sequence 
\bea 
0 \rightarrow V_+^{(p^2)} \rightarrow \widetilde V_{+} \rightarrow V_+ 
= \widetilde V_+ / V_+^{(p^2) }   \rightarrow 0 
\eea

The representation $ V_+$ is generated by acting with derivatives on the elementary scalar field, and using  the operator-state 
correspondence.  The representation $ V_+^{(p^2)}  $ is isomorphic to the representation obtained by taking 
all derivatives of $ \phi^3 $ in free scalar field theory and applying the operator-state correspondence.  
When we perturb the free theory with a $\phi^4$ interaction, we have the quantum equation of motion 
\bea 
\partial_{\mu} \partial_{ \mu} \phi = g \phi^3 
\eea
This quantum equation of motion, and its relation to the indecomposable representation $ \widetilde V_+$, 
 is reflected in the properties of the integral (\ref{IntegralOfInterest}). 
This will be the subject of Section \ref{sec:QEOM-indec}.  
Indecomposable representations have appeared in discussions of 2D CFT, see for example 
\cite{Gaberdiel0105,VJS11,DoFlo08}.
Our observations draw some elements from this work e.g. in the use we make of 
twisted co-products in connection with  OPEs, but they are not a direct translation of the 2D story 
which relies on the use of the complex coodinates $ (z , \bar z ) $ and the corresponding chiral-anti-chiral factorization. 

\section{Harmonic expansion method and the logarithmic term } 
\label{sec:HPEM}

The harmonic expansion method expands the two point function in terms of products of spherical harmonics.
In this way the action of $so(4,2)$ on any of the four external coordinates is manifest. 
The form of the expansion is dictated by the relative sizes of the integration variable and the external coordinates.
Consequently, this expansion method breaks the integration region down into a set of 5 regions.
The main result of this section is an explicit answer for each of these regions. This  allows us to 
isolate the logarithmic term to be discussed further in Section \ref{sec:CoeffLog}. It also  gives a neat separation 
 of the integral into terms which are  homogeneous and inhomogeneous terms for each of the Laplacians $ \Box_i$, 
  which will be useful for the equivariant interpretation of the quantum equations of motion 
  in section \ref{sec:QEOM-indec}. 

Let $ |x_1| < |x_2| < |x_3| < |x_4| $. 
First consider  the region where  $ |x| $ is  less than all the $|x_i|$. 
\bea\label{theI1integra} 
I_{1}  = \sum_{ l , l_i , m_i }  { \prod_{ i=1}^{4 }  Y^{ l_i}_{m_i} ( x_i' )   \over |x_1|^2 |x_2|^2 |x_3|^2 |x_4|^2} 
 \int_{ 0}^{|x_1 |}  dr r^3 
r^{ l_1 + l_2 +l_3 + l_4 } \int d^3 \hat x  \prod_{i=1}^4  Y_{l_i}^{m_i} ( \hat x )  
\eea
where $d^3 \hat{x}={1\over 2\pi^2}dS^3$ with $dS^3$ the standard measure on the unit sphere.
The last factor is a group theoretic factor which will appear in all of the five integration regions. 
We can write it as 
\bea\label{theCfac} 
C^{ m_1 ,m_2 , m_3 , m_4 }_{ l_1 , l_2  , l_3 , l_4 } 
= \sum_{ l_5 , m_5 , m_6  } C^{m_1 , m_2 ; l_5 }_{ l_1 , l_2 ; m_5 } { 1 \over ( l_5+1 ) } 
 C^{m_3 , m_4 ; l_5  }_{ l_3 , l_4 ; m_6 } \delta ( m_1 + m_2 , m_5 ) \delta ( m_3 + m_4 , m_6 ) g^{ m_5 , m_6 } 
\eea
where 
\bea\label{strucCons} 
Y_{ l_1 }^{m_1} Y_{ l_2}^{m_2} = \sum_{ l_5 , m_5 } C_{ l_1 , l_2 ;  m_5 }^{ m_1 , m_2  ; l_5} Y_{ l_5}^{m_5} 
\eea
This is the Clebsch-Gordan coefficient for multiplication of spherical harmonics on $S^3$. 
Selection rules for $ C^{ l_1 , l_2 ;  l_5 }_{ m_1 , m_2  ; m_5}$  imply that 
\bea 
  max ( l_1 , l_2 ) -  min ( l_1 , l_2 ) \le l_5  \le l_1 + l_2 
\eea
or, equivalently 
\bea 
| l_1 - l_2 |  \le l_5 \le l_1 + l_2 
\eea
If we multiply two symmetric traceless tensors $T_1 , T_2 $  of ranks $ l_1 $ and $l_2$, 
we can get something symmetric and traceless of rank $ l_1  + l_2$. If we contract 
two indices, one from each, we can reduce the rank by $2$. Further such contractions 
reduce the rank by multiples of $2$. And the maximum number of contractions is $ min ( l_1 , l_2 )$. 
For the 4-point coupling of spherical harmonics to be non zero, we need 
\bea 
l_4 \in \{ l_1 + l_2 + l_3 , l_1 + l_2 + l_3 - 2 , l_1 + l_2 + l_3 - 4 , \cdots \} 
\eea
Alternatively, a convenient way to parametrize the possibilities is given by 
\bea 
l_1 + l_2 - 2 k_{12}  = l_3 + l_4 - 2 k_{34} 
\eea
where $ 0 \le k_{12} \le min ( l_1 , l_2 ) $ and $ 0 \le k_{34} \le min ( l_3 , l_4 )  $. 

After doing  the integral 
\bea\label{I1eq} 
I_1  && =\sum_{  l_i , m_i }  { \prod_i   Y^{l_i}_{m_i} (  x_i'  )  \over |x_1|^2 |x_2|^2 |x_3|^2 |x_4|^2} 
 C^{ m_1 ,m_2 , m_3 , m_4 }_{ l_1 , l_2  , l_3 , l_4 }  
{ |x_1|^{ 4 + l_1 + l_2 + l_3 + l_4  } \over 4 + l_1 + l_2 + l_3 + l_4 }  
\eea
We will write $ I_1 = I_{1;1}^S$, which indicates that the radial position of the interaction point
coincides with the radial position of $ |x| $ as we evaluate this integral. The superscript indicates 
that the answer is a power series in the $ x_i'^{ \mu} $. 

Next consider the region $ |x_1| < |x| < |x_2| $. The contribution to the integral from this region is 
\bea\label{I2eq}  
&& I_{2} = \sum_{ l_i , m_i } {  Y^{l_1}_{m_1} (  x_1 ) \prod_{ i=2} Y^{l_i}_{m_i} ( x_i')   \over |x_2|^2 |x_3|^2 |x_4|^2 } 
\int_{|x_1| }^{ |x_2| }  dr~ r^3 ~     r^{-2} \int d^3 \hat x Y_{ l_1 }^{m_1} ( x') Y_{l_2}^{m_2} ( x ) 
Y_{l_3}^{m_3} ( x ) Y_{l_4}^{m_4} ( x )    \cr 
&& = \sum_{ l_i , m_i } 
{  Y^{l_1}_{m_1} (  x_1 ) \prod_{ i=2} Y^{l_i}_{m_i} ( x_i') \over |x_2|^2 |x_3|^2 |x_4|^2 } 
\int_{|x_1| }^{ |x_2| }  dr~ r^3 ~     r^{-2} r^{ -l_1 + l_2 + l_3 + l_4  } C_{l_1 , l_2 , l_3 , l_4 }^{ m_1 , m_2 , m_3 , m_4 }    \cr 
&& = \sum_{ l_i , m_i } Y^{ l_1}_{ m_1} ( x_1 ) { Y^{ l_2}_{ m_2} ( x_2' ) \over |x_2|^2 } 
{ Y^{ l_3}_{ m_3} ( x_3' ) \over |x_3|^2 }  { Y^{ l_4}_{ m_4} ( x_4' ) \over |x_4|^2 } 
 {  C_{l_1 , l_2 , l_3 , l_4 }^{ m_1 , m_2 , m_3 , m_4 }    \over ( - l_1 + l_2 + l_3 + l_4 +2 ) }
  \left (  r_2^{ - l_1 + l_2 + l_3 + l_4 + 2 } - r_1^{ - l_1 + l_2 + l_3 + l_4 +2 } \right )  \cr 
&& 
\eea 
We used  $ \hat x = \hat x'$, $ Y^{l}_{m} ( x')=r^{- l} Y^{l}_{m} ( \hat x ) $ and $ Y^l_m (x)= r^{l} Y^{l}_{m}(\hat x)$.
Note that $ - l_1 + l_2 + l_3 + l_4 \ge 0$ follows from the  selection rules for $ su(2)$ tensor products. 
Define $I_{2;1}^{S},I_{2;2}^{S}$ 
\bea 
&& I_{ 2;1}^S  =\sum_{ l_i , m_i } Y^{ l_1}_{ m_1} ( x_1 ) { Y^{ l_2}_{ m_2} ( x_2' ) \over |x_2|^2 } 
{ Y^{ l_3}_{ m_3} ( x_3' ) \over |x_3|^2 }  { Y^{ l_4}_{ m_4} ( x_4' ) \over |x_4|^2 } 
 {  C_{l_1 , l_2 , l_3 , l_4 }^{ m_1 , m_2 , m_3 , m_4 }    \over ( - l_1 + l_2 + l_3 + l_4 +2 ) } \left ( 
 - r_1^{ - l_1 + l_2 + l_3 + l_4 +2 } \right ) \cr
&& I_{2;2}^S = \sum_{ l_i , m_i } Y^{ l_1}_{ m_1} ( x_1 ) { Y^{ l_2}_{ m_2} ( x_2' ) \over |x_2|^2 } 
{ Y^{ l_3}_{ m_3} ( x_3' ) \over |x_3|^2 }  { Y^{ l_4}_{ m_4} ( x_4' ) \over |x_4|^2 } 
 {  C_{l_1 , l_2 , l_3 , l_4 }^{ m_1 , m_2 , m_3 , m_4 }    \over ( - l_1 + l_2 + l_3 + l_4 +2 ) } \left ( 
r_2^{ - l_1 + l_2 + l_3 + l_4 + 2 }\right ) \cr 
&& 
\eea
$I_{2;1}^S $ is obtained from the limit where the radial position of the integrated  interaction 
point coincides with the radial position of the external leg $ x_1$, i.e. where $ |x| = |x_1|$. 
The superscript indicates that this is a power series in the $ x_1^{ \mu}, x_2'^{ \mu } , x_3'^{\mu} , x_4'^{\mu} $ variables. 
$I_{2;2}^S$ is analogously defined in terms of $ |x| = |x_2|$. We have 
\bea 
I_2 = I_{2;1}^S + I_{2;1}^R 
\eea

Now consider the third region where $ |x| $ is in the middle
\bea \label{I3eq}
I_3 &&= \sum_{ l_i , m_i } { Y^{l_1}_{m_1} (x_1) Y^{l_2}_{m_2} (x_2) Y^{l_3}_{m_3} ( x_3') Y^{l_4}_{m_4} ( x_4') \over |x_3|^2 |x_4|^2}  C_{l_1 , l_2 , l_3 , l_4 }^{ m_1 , m_2 , m_3 , m_4 }  
\int_{ |x_2| }^{ | x_3| } dr ~ r^{3 - 4 + l_3 +l_4 - l_1 - l_2 } \cr 
&& =\sum_{ l_i , m_i }   { Y^{l_1}_{m_1} (x_1) Y^{l_2}_{m_2} (x_2) Y^{l_3}_{m_3} ( x_3') Y^{l_4}_{m_4} ( x_4')  \over |x_3|^2 |x_4|^2 }  C_{l_1 , l_2 , l_3 , l_4 }^{ m_1 , m_2 , m_3 , m_4 }  \log \left ( { |x_3| \over  |x_2| } \right ) \delta ( l_1 + l_2 , l_3 + l_4 ) \cr 
&& + \sum_{ l_i , m_i ; l \ne 0 }    { Y^{l_1}_{m_1} (x_1) Y^{l_2}_{m_2} (x_2) Y^{l_3}_{m_3} ( x_3') Y^{l_4}_{m_4} ( x_4')  \over |x_3|^2 |x_4|^2 }
\left  (   { |x_3|^{ l } - |x_2|^{ l } \over l }   \right ) 
\delta (l , - l_1 -l_2 +  l_3 + l_4 ) \cr 
&& 
\eea
It is convenient to define 
\bea 
&& I_{3;2}^{L} =  - \sum_{ l_i , m_i }   { Y^{l_1}_{m_1} (x_1) Y^{l_2}_{m_2} (x_2) Y^{l_3}_{m_3} ( x_3') Y^{l_4}_{m_4} ( x_4')  \over |x_3|^2 |x_4|^2 }  C_{l_1 , l_2 , l_3 , l_4 }^{ m_1 , m_2 , m_3 , m_4 }  \log \left (   |x_2| \right ) \delta ( l_1 + l_2 , l_3 + l_4 )  \cr 
&& I_{3;2}^{S} =   \sum_{ l_i , m_i ; l \ne 0 }    { Y^{l_1}_{m_1} (x_1) Y^{l_2}_{m_2} (x_2) Y^{l_3}_{m_3} ( x_3') Y^{l_4}_{m_4} ( x_4')  \over |x_3|^2 |x_4|^2 }C_{l_1 , l_2 , l_3 , l_4 }^{ m_1 , m_2 , m_3 , m_4 } 
\left  (   { - |x_2|^{ l } \over l }   \right ) 
\delta (l , - l_1 -l_2 +  l_3 + l_4 )
\cr 
&& I_{3;3}^{L } = \sum_{ l_i , m_i }   { Y^{l_1}_{m_1} (x_1) Y^{l_2}_{m_2} (x_2) Y^{l_3}_{m_3} ( x_3') Y^{l_4}_{m_4} ( x_4')  \over |x_3|^2 |x_4|^2 }  C_{l_1 , l_2 , l_3 , l_4 }^{ m_1 , m_2 , m_3 , m_4 }  \log \left (  |x_3| \right ) \delta ( l_1 + l_2 , l_3 + l_4 )  \cr 
&& I_{3;3}^S =  \sum_{ l_i , m_i ; l \ne 0 }    { Y^{l_1}_{m_1} (x_1) Y^{l_2}_{m_2} (x_2) Y^{l_3}_{m_3} ( x_3') Y^{l_4}_{m_4} ( x_4')  \over |x_3|^2 |x_4|^2 }C_{l_1 , l_2 , l_3 , l_4 }^{ m_1 , m_2 , m_3 , m_4 } 
\left  (   { |x_3|^{ l } \over l }   \right ) 
\delta (l , - l_1 -l_2 +  l_3 + l_4 ) \cr 
&& 
\eea
$I_{3;2}^L $ is the logarithmic term coming from the $|x| = |x_2|$ end of the integral, where the radial position of the interaction point coincides with the radius of the external point  $x_2$. 
$ I_{3;2}^S$ is the series term from the same limit. $ I_{3;3}^L , I_{3;3}^S$ have been defined analogously. Again,
\bea 
I_{3} = I_{3;2}^L + I_{3;2}^S + I_{3;3}^L + I_{3;3}^S 
\eea

The integer $ l$ appearing in (\ref{I3eq}) can be positive or negative. 
If we assume $ x_1  ,x_2 $ are small and $ x_3  , x_4 $ large - we can specialize the 
known answers to the integral. In that case, we know that
\bea 
&& |x_2| = max ( r_1 , r_2 ) = { 1 \over 2 } \left ( r_1 + r_2 + |r_1 - r_2| \right) \cr 
&& |x_3| = min ( r_3 , r_4 ) =  { 1 \over 2 } \left ( r_3 + r_4 - |r_3 - r_4| \right )
\eea
It is also useful to express the result in terms of $r_3' = 1/r_3$, which gives
\bea 
\log \left ( r_2 / r_3 \right ) = \log \left (  {( r_1 + r_2 + |r_1 - r_2| ) ( r_3' + r_4' + |r_3' - r_4'| ) } \right ) 
\eea
Note that in the limit of $ r_1 , r_2\sim \epsilon \rightarrow 0$, $ r_3' , r_4' \sim \epsilon \rightarrow 0$, 
this goes like $\log ( \epsilon^2 ) $ just like $ \log s$.
In Section \ref{sec:CoeffLog} we will give the precise relation between the coefficient of $\log s$ in the exact answer
(what we call $F_1(s,t)$) and the coefficient of $\log \left( r_2 / r_3 \right)$ computed above.

In the fourth region $|x|$ is between $|x_3|$ and $|x_4|$. 
\bea 
&& I_{ 4} =\sum_{ l_i , m_i } { Y^{l_4}_{m_4} (x_4' ) \prod_{i=1}^3  Y^{l_i}_{m_i} ( x_i )    \over |x_4|^2 }
 C^{l_1 , l_2 , l_3 , l_4 }_{ m_1 , m_2 , m_3 , m_4 } \int_{ |x_3|}^{ |x_4| }  dr ~ r^{ 3 - 6}   r^{  -l_1 - l_2 -  l_3 + l_4 }   \cr 
&& = \sum_{ l_i , m_i ; l \le 0  }
 { Y^{l_4}_{m_4} (x_4' ) \prod_{i=1}^3  Y^{l_i}_{m_i} ( x_i )    \over |x_4|^2 } 
\left  (   { |x_4|^{ l - 2  } - |x_3|^{ l -2 } \over l -2  }   \right ) \delta (l , - l_1 -l_2 -   l_3 + l_4 ) 
\eea
Note that  there are no log terms here since $ l$ is never equal to $2$. 
It is useful to define 
\bea 
&& I_{4;3} = \sum_{ l_i , m_i ; l \le 0  }
 { Y^{l_4}_{m_4} (x_4' ) \prod_{i=1}^3  Y^{l_i}_{m_i} ( x_i )    \over |x_4|^2 }
\left ( { - |x_3|^{ l -2 } \over l -2  }  \right ) 
\delta (l , - l_1 -l_2 -   l_3 + l_4 )  \cr 
&& I_{4 ; 4 } =  \sum_{ l_i , m_i ; l \le 0  }
 { Y^{l_4}_{m_4} (x_4' ) \prod_{i=1}^3  Y^{l_i}_{m_i} ( x_i )    \over |x_4|^2 }
\left  (   { |x_4|^{ l - 2  }  \over l -2  }   \right )
\eea
$I_{4;3} $ is obtained from the lower limit where $ |x| = |x_3|$, with the radial position of 
the interaction point  coinciding with the radial position of $x_3$.  $I_{4;4}$ is obtained from the upper limit $ |x| = |x_4| $ and
\bea 
I_{4 } = I_{4;3} + I_{4;4} 
\eea

The fifth region is given by $ |x| > |x_i| $. 
\bea 
&& I_5 =  \sum_{ l_i , m_i } C_{l_1 , l_2 , l_3 , l_4 }^{ m_1 , m_2 , m_3 , m_4 }   \prod_i Y^{l_i}_{m_i} (  x_i )  \int_{ |x_4| }^{ \infty}  dr r^{ 3 - 8 } r^{  -l_1 - l_2 - l_3 - l_4 }   \cr 
&& =     \sum_{ l_i , m_i ; l }  C_{l_1 , l_2 , l_3 , l_4 }^{ m_1 , m_2 , m_3 , m_4 }   \prod_i Y^{l_i}_{m_i} (  x_i )  \left  (   { |x_4|^{ -4 -  l  } \over  4 + l   }   \right ) \delta (l ,  l_1 + l_2 +  l_3 + l_4 )  
\eea
In this case there is no log term as the $l_i$ are all integers greater than or equal to zero. We write $I_5=I_{5;4}^S$ 
to indicate that this is a power series expansion and arises from the integral at the limit $ |x| = |x_4|$.  

The integral $I=I_1+I_2+I_3+I_4+I+5$ is a contribution to the four point function of free scalar fields, at points
$x_1,x_2,x_3,x_4$. Each field has dimension $\Delta=1$ and spin zero. Consequently, acting with the 
quadratic Casimir on any field must give
\bea
   C_2=\Delta (\Delta - 4)+l(l+2)=-3
\eea
In Appendix \ref{basicf} we explain how to translate $C_2$ into a differential operator.
Using the resulting differential operator $(C_2)_i$ in any of the coordinates $x_i$, we verify that
\bea\label{casimirsij}
   (C_2)_i I^{(j)}  =-3I^{(j)} \qquad i, j  \in \{ 1,2,3,4 \} 
\eea

\section{ Coefficient of the log term and the projector } 
\label{sec:CoeffLog} 

We are computing $I(x_1,x_2,x_3,x_4)$ with specified ordering  $|x_1|<|x_2|<|x_3|<|x_4|$.
Applying the HPEM, there is a logarithmic term coming from the range $|x_2 |\le  x\le |x_3|$.  
In this section we want to argue that the coefficient of the logarithmic term has a representation theory interpretation
as an invariant map built from a projection operator $ \cP_{++ ; ++ }$ that we define below.
The projection operator $ \cP_{++ ; ++ }$ featured prominently in the work of Frenkel and Libine\cite{FL07}.

The logarithmic term coming from the HPEM was computed in the last section. The result is
\bea\label{coefflog} 
\log ( { r_3 \over r_2 }  ) 
\sum_{ l_i , m_i } 
\left (   { Y^{l_1}_{m_1} (x_1) Y^{l_2}_{m_2} (x_2) Y^{l_3}_{m_3} ( x_3') Y^{l_4}_{m_4} ( x_4')  
     \over |x_3|^2 |x_4|^2 } 
 C_{l_1 , l_2 , l_3 , l_4 }^{ m_1 , m_2 , m_3 , m_4 }   \right )
 \delta ( l_1 + l_2 , l_3 + l_4 ) 
\eea
The exact result for $I$ was given in (\ref{exactresult}) in terms of
\bea 
\Phi ( s ,t ) = F_0 ( s , t ) + \log ( s ) F_1 ( s , t ) 
\eea
Consider the Casimir 
\bea 
C_2 =  - { 1 \over 2 }  \eta^{ AC } \eta^{BD} ( \ccL_{ AB }^{(1)} + \ccL_{AB}^{(2)} ) ( \ccL_{ CD }^{(1)} + \ccL_{CD}^{(2)} ) 
\eea
of $ so(4,2)$ acting on the coordinates $ x_1 , x_2$. For any function $H(s,t)$ of the conformal cross ratios the quadratic Casimir $C_2$ of $so(4,2)$ becomes
the differential operator\cite{DOcas}
\bea
C_2 H =
2(1+s-t)s t {\partial^2 H\over\partial s\partial t}
-\big(1-s+t\big)s
{\partial\over\partial s}\Big(s {\partial H\over\partial s}\Big)\cr
 -\Big( (1-t)^2-s(1+t)\Big){\partial\over\partial t}\Big( t{\partial H\over\partial t}\Big)+4s {\partial H\over\partial s}
\eea
Using the above differential operator, we find
\bea
C_2 \,\, s \Phi = - 4\,\, s\Phi\qquad\qquad\qquad\qquad\qquad\qquad C_2 \,\, sF_1  = - 4 \,\, sF_1
\eea
Thus the Casimir equation obeyed by the full integral is also obeyed  by the coefficient of the log term.
From (\ref{exactresult}) we see that the coefficient of $\log (s)$ in the known exact answer for the integral is
\bea
{1\over 2 x_{13}^2 x_{24}^2}F_1 ( s , t )
\eea
The $\log (s)$ appearing in (\ref{exactresult}) is the only possible source of $\log {r_3\over r_2}$ dependence,
which implies that
\bea
2\sum_{ l_i , m_i } 
\left (   { Y^{l_1}_{m_1} (x_1) Y^{l_2}_{m_2} (x_2) Y^{l_3}_{m_3} ( x_3') Y^{l_4}_{m_4} ( x_4')  
     \over |x_3|^2 |x_4|^2 } 
 C_{l_1 , l_2 , l_3 , l_4 }^{ m_1 , m_2 , m_3 , m_4 }   \right )
 \delta ( l_1 + l_2 , l_3 + l_4 ) =
{1\over x_{13}^2 x_{24}^2}F_1 ( s , t )
\eea

The representation $V_+$ has lowest weight state of dimension $1$, written as $V_+$. 
In the notation of Dolan \cite{Dolan2005}  it is $D_{ [1,0,0]} $.  The tensor product $ V_+ \otimes V_+$ can be decomposed 
into a direct sum of irreducible representations \cite{Dolan2005,Heiden1980}
\bea 
V_+ \otimes V_+  = \cA_{[200]}\, +\,  \bigoplus_{ k=1}^{ \infty} D_{ [ k+2 ,  { k \over 2 } , { k \over 2 } ]  }  
\eea
Given such a decomposition of a tensor product into a direct sum, there are projectors for each of the terms. 
These projectors commute with the $so(4,2)$ actions and hence describe equivariant maps. 
 The representation $ \cA_{ [2,0,0]}  $ will henceforth be called $V_{++}$ and corresponds to the CFT primary  operator  $ \phi^2$ and its descendants. 
There are Clebsch-Gordan maps 
\bea
\cM :&  V_+ \otimes V_+ & \rightarrow ~~V_{++} \cr 
\cM^{ \dagger} :&  V_{ ++} & \rightarrow ~~  V_+ \otimes V_+ 
\eea
which are equivariant maps between the tensor product and the irrep. 
There is a projector $\cP_{++;++}$ defined by
\bea\label{projp4} 
  \cP_{++ ; ++ } : V_+ \otimes V_+ \rightarrow V_+ \otimes V_+ \cr 
\cP_{ ++ ;  ++  } = \cM \circ \cM^{ \dagger} 
\eea
There is a closely related projector $ \cP_{ ++--} $ 
\bea 
\cP_{ ++--} : V_+ \otimes V_+ \otimes V_- \otimes V_- \rightarrow \mC 
\eea
This is obtained by tensoring both sides of (\ref{projp4}) with $ V_- \otimes V_-$, 
\bea\label{extend}  
&& \tilde  \cP : V_+ \otimes V_+ \otimes V_-  \otimes V_- \rightarrow V_+ \otimes V_+ \otimes V_-  \otimes V_- \cr 
&& \tilde \cP = \cP \circ 1_{V_- \otimes V_-}  
\eea
\begin{figure}[ht]%
\begin{center}
\includegraphics[width=0.9\columnwidth]{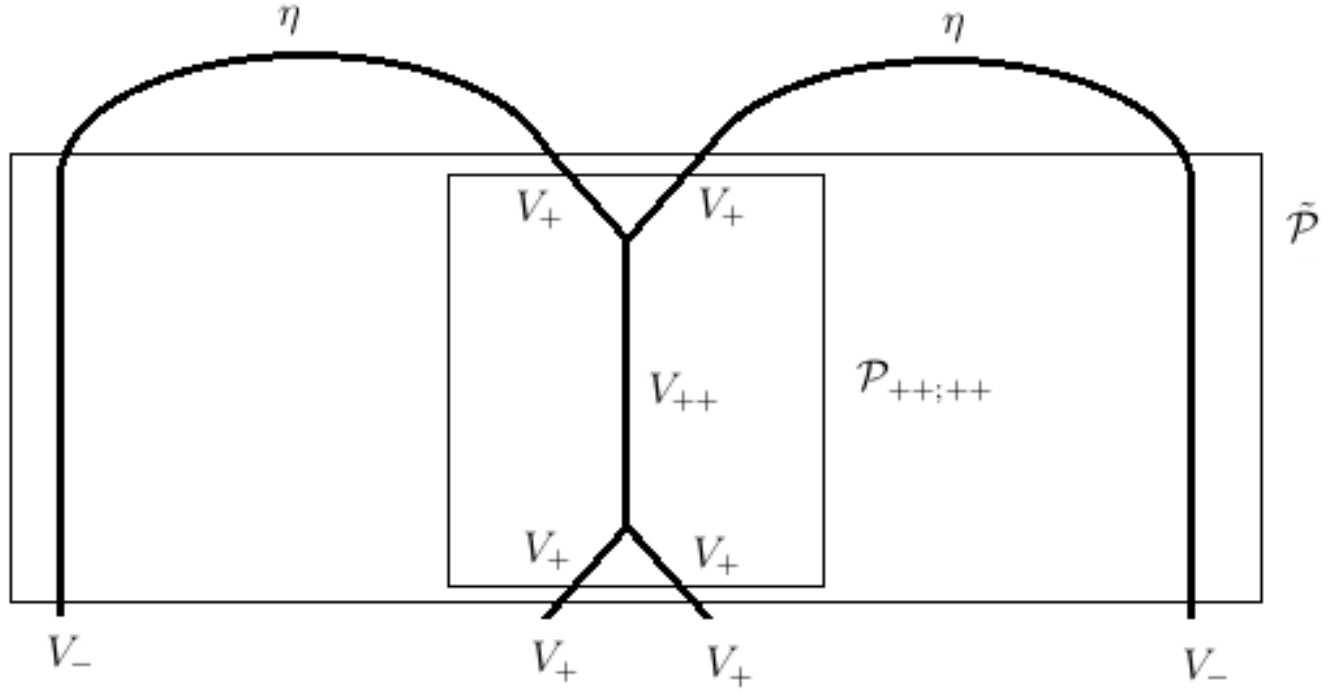}%
\caption{The construction of $\cP_{ ++--}$, follows by composing $\cP$ and two invariant pairings ($\eta$).}%
\label{fig:tildeP}%
\end{center}
\end{figure}
The RHS of the first line of (\ref{extend}) can be equivariantly mapped to $ \mC $ by using the invariant pairing 
 between the first $V_+$ and first $V_-$ and the invariant pairing between second $ V_+ $ and second $V_-$. 
Composing  $ \tilde\cP$ with these invariant pairings gives $\cP_{++;--}$. 
We can evaluate this projector on position eigenstates 
\bea\label{projOnPos}  
\cP_{ ++--} (e^{-iP\cdot x_1} v^+ \otimes e^{-iP\cdot x_2 } v^+ \otimes e^{iK\cdot x_3'} v^- \otimes 
e^{iK\cdot x_4'} v^- ) 
\equiv \cP ( x_1 , x_2 , x_3' , x_4' ) 
\eea
We can also evaluate it on spherical harmonics 
\bea 
\cP_{ ++--} ( Y^{l_1}_{m_1} ( P ) v^+ \otimes Y^{ l_2 }_{m_2} ( P ) v^+ \otimes Y^{l_3}_{m_3} ( K ) v^- \otimes 
Y^{l_4}_{m_4} ( K ) v^-  ) \equiv \cP^{ l_1 , l_2 , l_3 , l_4}_{ m_1 , m_2 , m_3 , m_4 }  
\eea

Our claim is that the power series expansion of ${1\over 2 x_{13}^2 x_{24}^2} F_1 $ at small $ x_1 , x_2 , x_3' , x_4' $ coincides 
with that of $ \cP (x_1,x_2,x_3',x_4')$ 
\begin{equation}\label{theclaim}  
\boxed{ 
~~~~~{1\over 2 x_{13}^2 x_{24}^2} F_1 ( u ( x_1 , x_2 , x_3' , x_4'  ) , s ( x_1 , x_2 , x_3' , x_4'  )  ) = x_3^{\prime 2} x_4^{\prime 2}
\cP ( x_1 , x_2 , x_3' , x_4' )  ~~~~~~
}
\end{equation} 
 This is the main result of this section. 
This power series expansion can be conveniently organised in terms of the coefficients 
$ \cP^{ l_1 , l_2 , l_3 , l_4}_{ m_1 , m_2 , m_3 , m_4 }  $. 

Consider the coefficient of the product of harmonics in the log term (\ref{coefflog}).  
The spherical harmonics $Y^l_m(x)$ are a basis for harmonic functions regular at $x=0$ and carry a representation of $so(4,2)$.
Denote the corresponding function space $H_+$. 
Similarly, $(x')^2 Y^l_m(x')$ are a basis for harmonic functions regular at $x'=0$, i.e. $x=\infty$.
They also carry a representation of $so(4,2)$.
Denote the corresponding function space $H_-$. 
Picking up the coefficient of the harmonics is mapping $H_+ \otimes H_+ \otimes H_- \otimes H_- \rightarrow  \mC$.
This coefficient is just the tensor $ C_{l_1,l_2,l_3,l_4}^{m_1,m_2 , m_3 , m_4}$ 
 defined in equations (\ref{theCfac}) and (\ref{strucCons}) in terms of the structure constants for multiplication of spherical harmonics. 
The 3-point  structure constants involve the  integration
\bea 
\int { d^3 \hat x } ~~ Y^{ l_1 }_{ m_1 } ( \hat x ) Y^{ l_2 }_{m_2} ( \hat x ) Y^{ l_5}_{ m_5} ( \hat x )  
\eea
Thus picking up the coefficient of $ Y^{ l_1 }_{ m_1 } ( x_1  ) Y^{ l_2 }_{m_2} ( x_2  ) $ 
involves mapping 
\bea\label{themapM}  
&& Y^{ l_1 }_{ m_1 } (x_1) \otimes Y^{ l_2 }_{m_2} (x_2) \rightarrow r^{l_1+l_2}Y^{ l_1 }_{m_1}(\hat x)Y^{ l_2 }_{m_2}(\hat x ) \cr 
&&  { Y^{ l_3 }_{ m_3 } (x_3') \over |x_3|^2 } \otimes { Y^{ l_4 }_{m_4} (x_4') \over |x_4|^2 } 
 \rightarrow { Y^{ l_3 }_{ m_3 } (x') \over |x|^2 } \otimes { Y^{ l_4 }_{m_4} (x') \over |x|^2 }
\eea 
These are applications of the equivariant maps 
 $ \cM : H_+ \otimes H_+ \rightarrow H_{ ++}$ and $ \cM : H_- \otimes H_- \rightarrow H_{--}  $ 
  as explained in  section 5.2 of \cite{FL07}. From a physical perspective, this corresponds to the fact that the free scalar field $ \phi(x) $ has modes transforming in $ H_+ $ (and $H_-$)  
 while the field $ \phi^2 (x)$ has modes transforming in $ H_{++} $ ( and $H_{--}$). 
After these maps are applied, the HPEM sets $l_1 + l_2 = l+3 + l_4 $ when we do the radial 
integral and pick up the log term. There remains an integral over $S^3$ which gives the factor $ (l_5+1)^{-1} $. This corresponds, in the discussion  of  \cite{FL07} (proposition 84) (see also equation (12) of \cite{Lib1309}),  to an integral over $U(2)$. Thus we have a direct link between the 
integration over a modified contour in complexified space-time (where we are integrating over 
$U(2)$ instead of MInkowski space) and the coefficient of  the log term. This is likely to be an example of a general story that should hold for more general Feynman integrals.

We can use the vertex operators of TFT2 to further clarify the discussion.
The natural language for the above discussion is in terms of a map 
$H_+ \otimes H_+ \otimes H_- \otimes H_- \rightarrow  \mC$.
Using the vertex operators we will see that it is equally natural to employ a map
$V_+ \otimes V_+ \otimes V_- \otimes V_- \rightarrow  \mC$.
Start with (\ref{projOnPos}) and expand the exponentials in $ Y^l_m(P)Y^m_l(x)$ or $Y^l_m(K)Y^m_l(x')$. 
\bea 
 && \sum_{ l_i , m_i } {( -i)^{ l_1} \over 2^{ l_1} l_1 ! }  Y^{ l_1}_{ m_1} ( x_1  )
 { (-i)^{ l_2} \over 2^{ l_2} l_2 ! }  Y^{ l_2}_{ m_2} ( x_2  ) 
{ (i)^{ l_3} \over 2^{ l_3} l_3! } Y^{ l_3 }_{ m_3} ( x_3' ) 
{ (i)^{ l_4} \over 2^{ l_4} l_4! } Y^{ l_4 }_{ m_4} ( x_4' ) \cr 
&& \cP_{ ++--}  \left (    Y_{ l_1}^{ m_1} ( P )      v^+ \otimes 
 Y_{ l_2}^{ m_2} ( P )      v^+    \otimes  Y_{ l_3}^{ m_3} ( K  )      v^-
\otimes Y_{ l_4}^{ m_4} ( K) v^-  
\right )  \cr 
&& =  \sum_{ l_i , m_i } Y^{ l_1}_{ m_1} ( x_1  )
  Y^{ l_2}_{ m_2} ( x_2  ) 
Y^{ l_3 }_{ m_3} ( x_3' ) Y^{ l_4 }_{ m_4} ( x_4' ) \cr 
&& \cP_{ ++--}  \left (    {  Y_{ l_1}^{ m_1} ( P ) \over 2^{ l_1} l_1 ! } 
      v^+ \otimes 
{  Y_{ l_2}^{ m_2} ( P ) \over 2^{ l_2} l_2! }       v^+    \otimes  
{ Y_{ l_3}^{ m_3} ( K  ) \over 2^{ l_3} l_3! }       v^-
\otimes {  Y_{ l_4}^{ m_4} ( K) \over 2^{ l_4} l_4! }  v^-  
\right ) 
\eea
The vertex operators of TFT2 provide equivariant maps between the algebraic state spaces and the polynomial state spaces 
which makes it possible to express the projector in terms of these state spaces. 
Indeed, the above argument makes it clear that $\prod Y^{ l_i}_{ m_i} (x_i )$  multiplies a  projector acting on states in 
$V_+\otimes V_+\otimes V_-\otimes V_- $.
This shows that the claim that the series expansion  multiplying the log is the evaluation of an $so(4,2)$
invariant projection on states created from the $v^{ \pm} $ by vertex operators, is equivalent to saying that the coefficient of the
product of spherical harmonics has to be an equivariant map. 

\subsection{Analytic consequences} 

We have seen that the coefficient of the log term in the HPEM has an expansion in powers of $ x_1 , x_2 , x_3' , x_4'$. 
We will now see how the same expansion arises from the exact answer.  

We want to consider the limit $s \rightarrow 0 , t \rightarrow 1 $, for the conformal cross ratios.
It proves to be useful to set $ t = 1 + u$ and then consider $ u \rightarrow 0 $. 
The coefficient of $\log s$  is
\bea 
F_1 = { 1 \over \lambda  } \log  \left ( ( 1 + \rho s ) \over ( 1 + \rho^{-1} t^{-1} ) \right ) 
\eea
We can expand this to low powers of $ s , u $ using Mathematica, and analysing the results, we arrive at 
\bea 
F_1 ( s , u ) = \sum_{ k=0}^{ \infty} 
{ s^k \over u^{ 2k+1} } [ Q_k (u ) \log ( 1 + u )   ]_{+ } 
\eea
Here $ Q_k(u) $ is a polynomial in $u$ and 
$ [ Q_k (u ) \log ( 1 + u )   ]_{+ } $ is the truncation of the power series in $u$ to powers $u^n$ with $n\ge 2k+1$.  
$Q_k(u)$ is order $k$
\bea 
Q_k (u) = \sum_{ l =0 }^k b_l(k) u^{ l} 
\eea
where 
\bea 
b_l ( k ) = { k \choose k-l } { 2k+l \choose k -l } 
\eea
Note that the existence of such an expansion is non-trivial. The individual 
factors such as $\rho , \lambda,$ when expanded in positive powers of 
$s$ contain, at each order, a finite number of negative powers of $u$. 
Nevertheless, the combination of terms appearing in $F_1$ 
is analytic in $ u , s $ at $ u , s \sim 0 $.  Appendix \ref{App:expExact} 
explains how we arrived at the above formula, with the help of Mathematica. 
The Appendix also explains how  the discussion implies 
a summation formula for products of $su(2)$ Clebsch-Gordan coefficients 
in terms of $F_1$.

\section{ Quantum Equation of motion, Indecomposable Representations and Equivariant maps }
\label{sec:QEOM-indec}

Using the harmonic expansion method, we have found
\bea 
&&  I = I_1 + I_2 + I_3 + I_4 + I_5 \cr 
&& I =  ( I_{1;1}^S + I_{2;1}^S )  + ( I_{2;2}^S + I_{3;2}^S + I_{3;2}^L) + 
( I_{3;3}^L + I_{3;3}^S + I_{4;3}^S) + ( I_{4;5}^S + I_{5;5}^S )  
\eea
Rearrange these contributions by defining 
\bea 
&& I^{(1)} = ( I_{1;1}^S + I_{2;1}^S )  \cr 
&& I^{(2)} =  ( I_{2;2}^S + I_{3;2}^S + I_{3;2}^L) \cr 
&& I^{(3)} = ( I_{3;3}^L + I_{3;3}^S + I_{4;3}^S)  \cr 
&& I^{(4)} = ( I_{4;5}^S + I_{5;5}^S )  
\eea
This reorganization is such that $I^{(i)}$ arises from integration limits where the radial position
of the interaction point coincides with the $i$'th external coordinate. 
In terms of the quantities just introduced, we have
\bea 
I = I^{(1)} + I^{(2)} + I^{(3)} + I^{(4)} 
\eea
Due to radial ordering, the order of the fields within the correlator swaps when moving from one term to the next.
As a consequence of these discontinuities we expect that
\begin{equation}\label{expectedqeom}
\boxed{
~~~~~~\Box_j I^{(i)} = -2 \delta_{i j } \prod_{ k \ne j  } { 1 \over ( x_k - x_i  )^2 } ~~~~~~
}
\end{equation}
We will  demonstrate, using the explicit formulae from the HPEM, that this is indeed the case.  
The $ I^{(i)}$ are also used to develop  equivariant map interpretations for 
the full integral $I$. Each $I^{(i)}$ is the starting point for one equivariant map 
interpretation. We exhibit the complete story for $ I^{(1)}$, while the discussion for $I^{(4)}$ 
 is related by inversion. We outline the story for $I^{(2)}$ (and by inversion for $I^{(3)}$). It has an additional intricacy involving the use of 
a twisted co-product. This raises some technical problems which we leave for the future. 

\subsection{ Quantum equations of motion } 

Consider the term $ I^{(1)}$, which is given by
\bea 
I^{(1)} = \sum_{ l_i , m_i } Y^{ l_1}_{ m_1} ( x_1 ) { Y^{ l_2 }_{ m_2 } ( x_2'  ) \over r_2^2}{  Y^{ l_3}_{ m_3} ( x_3'  ) \over r_3^2 } 
 {  Y^{ l_4}_{ m_4} ( x_4'  ) \over r_4^2 }  C_{ l_1 l_2 l_3 l_4}^{ m_1 m_2 m_3 m_4 }  
  { ( -2) ( l_1 + 1 ) r_1^{ 2 + l_2 + l_3 + l_4 - l_1 } \over ( l_1 + l_2 + l_3 + l_4 +4 ) ( - l_1 + l_2 + l_3 + l_4 + 2 ) }  \cr  
\eea
We have written  the above formula in terms of a product of harmonic functions in $ x_1 , x_2' , x_3' , x_4' $ so that it has a
smooth $ x_1 \rightarrow 0 $ limit as well as a smooth $ x_2 , x_3 , x_4\rightarrow\infty$  limit. 
To apply the Laplacian, to the above result the formulas
\bea \label{genIds}
&& { \partial^2  \over \partial x_1^{ \mu }  \partial x_1^{ \mu } } 
 ( r_1^2 )^{ A } = 4 A ( A +1 ) (  r_1^2 )^{ A -1} \cr
 &&{  \partial \over \partial x_1^{ \mu } } (  r_1^2 )^A  = 2 A ( r_1^2 )^{ A -1 } x_1^{ \mu } \cr  
&&   { \partial \over \partial x_1^{ \mu } } Y^{ l_1 }_{ m_1} ( x_1 )
 { \partial \over \partial x_1^{ \mu } }  ( r_1^2 )^{ A }   = 2 A l_1 ( r_1^2 )^{ A -1 } Y^{ l_1 }_{ m_1} ( x_1 )
\eea
are useful.
It is now simple to obtain
\bea\label{box1onI1}  
&& { \partial^2  \over \partial x_1^{ \mu }  \partial x_1^{ \mu } }  I^{(1)}
 = \sum_{ l_i , m_i  } { - 2 ( l_1 + 1 )  ( r_1  )^{ l_2 + l_3 + l_4 - l_1 +2 } \over r_2^2 r_3^2 r_4^2 }
Y^{ l_1}_{ m_1 }  ( x_1 ) Y^{ l_2}_{ m_2} (  x_2' ) Y^{ l_3 }_{ m_3 } ( x_3' ) Y^{ l_4}_{ m_4} (  x_4'  )
 C_{ l_1 l_2 l_3 l_4 }^{ m_1 m_2 m_3 m_4 } \cr
&&
\eea
To recognize the right hand side, note that
\bea 
{ 1 \over |x_1 - x_2 |^2 | x_1 - x_3  |^2 | x_1 - x_4|^2 }
&& = { 1 \over r_2^2 r_3^2 r_4^2 } \sum_{ l_i , m_i } Y^{ l_2}_{ m_2} ( x_1 ) Y^{ l_3}_{ m_3} ( x_1 ) Y^{ l_4}_{ m_4} ( x_1 ) 
Y^{ l_2}_{ m_2} ( x_2' ) Y^{ l_3}_{ m_3} ( x_3' ) Y^{ l_4}_{ m_4} ( x_4' )  \cr 
&& \!\!\!\!\!\!\!\!\!\!\!\!\!\!\!\!\!\!\!\!
= {( l_1 + 1 )r_1^{ l_2 + l_3 + l_4 } \over r_2^2 r_3^2 r_4^2 } \sum_{ l_i , m_i } C_{ l_1 l_2 l_3 l_4 }^{ m_1 m_2 m_3 m_4 }  
Y^{ l_1 }_{ m_1 } ( x ) Y^{ l_2}_{ m_2} ( x_2' ) Y^{ l_3}_{ m_3} ( x_3' ) Y^{ l_4}_{ m_4} ( x_4' )  
\eea
The $ (l_1+1) $ in the numerator arises because of the normalization of the spherical harmonics. 
Clearly then, we have demonstrated (\ref{expectedqeom}).

The harmonic expansion method expands the propagators in spherical harmonics which solve Laplace's equation. 
How then did we get a non-zero answer?  
The point is that when $ |x| < |x_1|$ we are expanding in positive powers of $x_1'$, and when $|x|>|x_1|$, we are expanding 
in positive powers of $ x_1$. 
In each case although the $x_1$ dependent functions are harmonics, the integration produces an additional dependence on $x_1$
from the integration limits.
In the operator formalism  where 
we compute a radially ordered correlator  the ordering of the interaction vertex changes relative to the external point $x_1$  when we move from the region $|x|<|x_1|$ to the region $|x|>|x_1|$.
So, as expected the violation of the free equation has to do with collision of the integration point with an external coordinate. 

The contribution $I^{(1)}$ did not include a log dependence.
We will consider one more example, $I^{(2)}$, chosen because this term does include a log dependence
\bea
I^{(2)}
&=& 
\sum_{l_i,m_i}Y^{l_1}_{m_1}(x_1)Y^{l_2}_{m_2}(x_2)Y^{l_3}_{m_3}(x_3')Y^{l_4}_{m_4}(x_4')
C_{l_1 l_2 l_3 l_4}^{m_1 m_2 m_3 m_4}{1\over l_2+l_3+l_4-l_1+2}
{\delta ( l_1 + l_2,   l_3 +  l_4 ) \over r_3^2 r_4^2} \cr 
&& + 
\sum_{l_i,m_i} Y^{l_1}_{m_1}(x_1) Y^{l_2}_{m_2}(x_2) Y^{l_3}_{m_3}(x_3') Y^{l_4}_{m_4}(x_4')
C_{l_1 l_2 l_3 l_4}^{m_1 m_2 m_3 m_4}{1\over r_3^2 r_4^2} \cr
&&\times\left[
-{r_2^{-l_1-l_2+l_3+l_4}2(l_2+1)\over (l_2+l_3+l_4-l_1+2)(l_3+l_4-l_1-l_2)} ( 1 - \delta (l_1+l_2,l_3+l_4) ) 
-\delta (l_1+l_2,l_3+l_4){\rm log} (r_2) 
\right] \nonumber
\eea
We will again make use of the formulas above in (\ref{genIds}) as well as
\bea
{\partial\over\partial x_2^{\mu}} Y^{l_2}_{m_2}(x_2)
 {\partial\over\partial x_2^{\mu }}(r_2^2)^A=2A l_2(r_2^2 )^{A-1}Y^{ l_2 }_{ m_2} ( x_2 )\cr
{\partial\over\partial x_2^\mu}{\rm log}r_2 = {x_2^\mu\over r_2^2}\qquad\qquad
{\partial\over\partial x_2^\mu}{\partial\over\partial x_2^\mu}{\rm log}r_2 = {2\over r_2^2}
\eea
We find
\bea\label{LonI2}
{\partial^2\over\partial x_2^\mu\partial x_2^\mu}I^{(2)}=
-2\sum_{l_i,m_i} (l_2+1){r_2^{-l_1-l_2+l_3+l_4-2}\over r_3^2 r_4^2}
Y^{l_1}_{m_1}(x_1)Y^{l_2}_{m_2}(x_2)Y^{l_3}_{m_3}(x_3')Y^{l_4}_{m_4}(x_4')
C_{l_1 l_2 l_3 l_4}^{m_1 m_2 m_3 m_4}
\eea
The right hand side can again be identified with
\bea
-{2\over |x_1 - x_2 |^2 | x_2 - x_3  |^2 | x_2 - x_4|^2 }
\eea
The log contributes the term with $l_1+l_2=l_3+l_4$ in (\ref{LonI2}).

The discussion for the terms $I^{(3)}$ and $I^{(4)}$ is now straight forward.

\subsection{ QEOM, equivariant maps and their lifts  } 

As we discussed there is an equivariant map between $ V_+^{ (p^2)} $ and 
the irrep generated by the field $ \phi^3 $, i.e. the irrep  $ V_{ +++} $ obtained 
by acting with $ ( P \cdots P ) $ on $ v^+ \otimes v^+ \otimes v^+ $. 
Given the TFT2 construction of free field correlators \cite{CFT4TFT2}, we know that there is an $so(4,2)$ equivariant map 
\bea 
{ \mathcal{F}}_1  : V_{+++ } \otimes V_- \otimes V_- \otimes V_- \rightarrow \mC 
\eea    
such that 
\bea 
{ \mathcal{F}}_1 ( e^{ -i P\cdot x_1 } v_{ +++} \otimes e^{ i K\cdot x_2' } v^-  \otimes  e^{ i K\cdot x_3' } v^-  \otimes 
e^{ i K\cdot x_4' } v^- ) = 
{ x_2^{ 2 } x_3^{ 2 } x_4^{2}  \over ( x_1 - x_2)^2 ( x_1 - x_3 )^2 ( x_1 - x_4 )^2 } 
\label{fmapexpected}
\eea
For completeness,we give a derivation   in Appendix \ref{appsec:QEOM-1}.  
Given the isomorphism between $ V_+^{ (p^2)} $ and $V_{+++}$, 
we have a map 
\bea 
{\mathcal{F}}_1 :  V_+^{(p^2)}   \otimes V_- \otimes V_-  \otimes V_- \rightarrow \mC 
\eea
It is given similarly by 
\bea\label{equivVp2} 
{\mathcal{F}}_1(e^{-i P\cdot x_1 }P_{ \mu }P_{ \mu}v^+ \otimes e^{iK\cdot x_2' }v^-\otimes e^{iK\cdot x_3'}v^-\otimes e^{iK\cdot x_4' }v^-) 
\!\!\!
&& = ~  {x_2^{2}x_3^{2}x_4^{2}\over (x_1-x_2)^2 (x_1-x_3 )^2 (x_1-x_4 )^2}\cr 
&&= ~ f ( x_1 , x_2' ) f ( x_1 , x_3' ) f ( x_1, x_4' )  \cr &&
\eea
The function $ f ( x_1 , x_2' ) $ is  the series in positive powers of $ x_1 , x_2' $ which sums to 
\bea 
{ 1 \over ( 1 - 2 x_1 \cdot x_2' + x_1^2 x_2'^2 ) } 
\eea

A consistency check of this interpretation is that the Casimirs for each of the four $so(4,2)$'s, 
one for each coordinate $x_i$, gives the value $(-3)$ appropriate for
  $ V_{ \pm } , \widetilde V_{\pm} $
  (\ref{casimirsij}). 

This map $ \cF_1$ can be lifted from the  subspace $ V_+^{(p^2)}  $ to the larger space $ \tilde V_+$ 
\bea 
\widetilde \cF_1 : \widetilde V_+  \otimes V_- \otimes V_- \otimes V_- \rightarrow \mC 
\eea 
Using the relation between algebraic generators $P_{\mu}$ and derivatives in the presence of the vertex operators, 
this  implies
\bea 
{ \partial^2  \over  \partial x_1^{ \mu} \partial x_1^{ \mu} }  
{\widetilde {\mathcal{F}}_1 } 
( e^{-iP\cdot x_1}v^+ \otimes e^{iK\cdot x_2'}v^- \otimes  e^{iK\cdot x_3'}v^-\otimes e^{iK\cdot  x_4'}v^-)= f (x_1,x_2') f (x_1,x_3') f(x_1,x_4') 
\eea
 $\widetilde { \mathcal F}_1$ is determined by ${ \mathcal F}_1$ up to   terms harmonic in $x_1$.
We know that $ x_2^2 x_3^2 x_4^2 I^{(1) } ( x_1 , x_2 , x_3 , x_4 )$ solves this differential equation, so we can identify 
\bea\label{equivVtilde}  
 x_2^2 x_3^2 x_4^2 ~~  I^{(1)} ( x_1 , x_2 , x_3 ,x_4 ) \rightarrow 
\widetilde {\mathcal{F}}_1 (e^{-i P\cdot x_1}v^+\otimes e^{iK\cdot x_2'}v^-\otimes e^{iK\cdot x_3'}v^-\otimes e^{iK\cdot x_4'}v^-)
\eea 
While $  x_3^2 x_4^2 I^{(1)}$ is an $so(4)$-equivariant map, it is  not $so(4,2)$ equivariant, even though the Laplacian in $x_1$  acting on it gives the $so(4,2)$ equivariant map $\cF_1$. 
The equivariance condition under the action of the momentum operator 
\bea 
&& \widetilde \cF_1 ( P_{ \mu} e^{ - i P\cdot x_1} v^+ \otimes e^{ iK\cdot x_2} v^- \otimes e^{ i K\cdot x_3} v^- \otimes e^{ i K\cdot x_4} v^- ) \cr
&&+  \widetilde \cF_1 (  e^{ - i P\cdot x_1} v^+ \otimes P_{\mu} e^{ iK\cdot x_2} v^- \otimes e^{ i K\cdot x_3} v^- \otimes e^{ i  K\cdot x_4} v^- ) \cr 
&& + \widetilde \cF_1 (  e^{ - i P\cdot x_1} v^+ \otimes e^{ iK\cdot x_2} v^- \otimes P_{\mu }  e^{ i K\cdot x_3} v^- \otimes e^{ i K\cdot x_4} v^- ) \cr
&&+ \widetilde \cF_1 (  e^{ - i P\cdot x_1} v^+ \otimes e^{ iK\cdot x_2} v^- \otimes e^{ i K\cdot x_3} v^- \otimes P_{\mu}  e^{ i  K\cdot x_4} v^- ) 
=0 
\eea
implies that 
\bea 
\left ( { \partial \over \partial x_1^{\mu} }  + x_2^2 { \partial \over \partial x_2^{\mu} }  { 1 \over x_2^2 } 
+ x_3^2 { \partial \over \partial x_3^{\mu} }  { 1 \over x_3^2 }  + x_4^2 { \partial \over \partial x_4^{\mu} }  { 1 \over x_4^2 } \right )  \widetilde \cF_1 = 0  
\eea
This condition is not satisfied if we identify $ \widetilde \cF_1 \rightarrow x_2^2 x_3^2 x_4^2 I^{(1)} $. 
We can add homogeneous terms, annihilated by $ \Box_1$ 
to get $  x_2^2 x_3^2 x_4^2 ( I^{(1)} + I^{(2)} + I^{(3) } + I^{(4)} ) = x_2^2  x_3^2 x_4^2 I $. 
Now equivariance under $ P_{\mu}$ action  of $\widetilde \cF_1$ follows from the standard translation invariance of the integral $ I$ 
\bea
\left ( { \partial \over \partial x_1^{\mu} } + { \partial \over \partial x_2^{\mu} } + { \partial \over \partial x_3^{\mu} }  + { \partial \over \partial x_4^{\mu} }  \right )   I  = 0 
\eea
Similar remarks hold for invariance under the special conformal transformations $ K_{\mu}$ and the dilatation operator $D$. 
Hence the quantum equation of motion along with  the requirement of  $so(4,2)$ equivariance condition identifies the lift $ \widetilde \cF_1$ as  
\begin{equation}\label{main2} 
\boxed{ 
\widetilde \cF_1 (  e^{ - i P\cdot x_1} v^+ \otimes e^{ iK\cdot x_2} v^- \otimes e^{ i K\cdot x_3} v^- \otimes e^{ i K\cdot x_4} v^- ) = x_2^2 x_3^2 x_4^2 I ( x_1 , x_2 , x_3 , x_4 ) = { x_2^2 x_3^2 x_4^2 \over x_{13}^2 x_{24}^2 } \Phi( s , t )
}
\end{equation} 

By inversion, a similar discussion holds for $I^{(4)}$ and the QEOM 
for $x_4$.
\bea 
\cF_4 : V_+ \otimes V_+ \otimes V_+ \otimes \tilde V_- \rightarrow \mC 
\eea
with 
\bea
 \cF_4(e^{-i P\cdot x_1}v^+\otimes e^{-iP\cdot x_2}v^+\otimes e^{-iP\cdot x_3}v^+\otimes e^{iK\cdot x_4'}K_{ \mu } K_{ \mu} v^-)
={x_1^{\prime 2} x_2^{\prime 2} x_3^{\prime 2} \over (x_1'-x_3')^2 (x_2'-x_3')^2 (x_3'-x_4')^2}  \cr
\eea
The $   x_1^2 x_2^2 x_3^2 I^{(4)}$ integral is an $so(4)$ invariant lift of $\cF_4$. 
\bea 
 \cF_4(e^{-i P\cdot x_1}v^+\otimes e^{-iP\cdot x_2}v^+\otimes e^{-iP\cdot x_3}v^+\otimes e^{iK\cdot x_4'}v^-)
  \rightarrow x_1^{\prime 2} x_2^{\prime 2} x_3^{\prime 2} ~~  I^{(4)} ( x_1 , x_2 , x_3 ,x_4 )
\eea 
The $so(4,2)$ equivariant lift is again given by adding homogeneous terms  
\bea 
\cF_4(e^{-i P\cdot x_1}v^+\otimes e^{-iP\cdot x_2}v^+\otimes e^{-iP\cdot x_3}v^+\otimes e^{iK\cdot x_4'}v^-)
= x_1'^2 x_2'^2 x_3'^2 I ( x_1 , x_2 , x_3 , x_4 ) 
\eea

\subsection{ QEOM and twisted equivariant map} 

In the above discussion the  solution  $I^{(1)}$ to the QEOM 
is  not logarithmic. Logarithmic contributions are added to ensure  $so(4,2)$ equivariance
of the lift from $ V_+^{(p^2)}$ to $ \widetilde V_+$. 
It is interesting to see how things are modified when 
we consider the case of $I^{(2)}  $, which is  a logarithmic solution to the quantum equation of motion. 
It is instructive to consider the free field  correlator 
\bea 
\langle \phi ( x_1 ) \phi^3 ( x_2 ) \phi ( x_3 ) \phi ( x_4 ) \rangle 
\eea
In the range $ |x_1| < |x_2| < |x_3| <|x_4| $ relevant to $I^{(2)}$, this free-field correlator,
with the correct series expansion, is constructed by taking 
$\phi^-(x_2')\otimes\phi^{+}(x_2)\otimes\phi^+(x_2)$ at $x_2$ and applying invariant pairings with 
  $\phi^+(x_1)$ and $\phi^- (x_3' )\otimes \phi^- ( x_4') $. In the free field  CFT4/TFT2 construction 
  we used  $ \phi = \phi^+ + \phi^- $ and the composite field $ \phi^{\otimes 3 } $ involved sums including 
  $ \phi^+\otimes\phi^+\otimes\phi^+$ and $\phi^-\otimes\phi^+\otimes\phi^+$. 
For such sums to be  $so(4,2)$ covariant we must use a twisted co-product.

There is a family of  automorphisms of $ so(4,2)$ parametrized by  a number $ \lambda $ 
\bea
\alpha_\lambda (P_\mu)={K_{\mu}\over\lambda}\qquad
\alpha_\lambda (K_\mu)=\lambda P_{\mu}\cr
\alpha_\lambda (M_{\mu\nu})=M_{\mu\nu}\qquad
\alpha_\lambda (D)=-D
\eea
A homomorphism between $ so(4,2)$ and $so(4,2)^{\otimes 4 } $ is given by the twisted co-product 
\bea 
\Delta_{\lambda} ( \ccL_a ) = \alpha_{ \lambda} (\ccL_a) \otimes 1 \otimes 1 \otimes 1 + 1 \otimes \ccL_a\otimes 1 \otimes 1 +  
1 \otimes 1 \otimes \ccL_a \otimes 1 +  1 \otimes 1 \otimes 1 \otimes \ccL_a
\eea
 We can write a new version of (\ref{equivVp2})
\bea
\cF_2 (e^{iP\cdot x_1}v^+\otimes e^{-iP\cdot x_2} ( P_{\mu} P_{\mu} v^{+}) \otimes e^{iK\cdot x_3'}v^-\otimes e^{iK\cdot x_4'}v^-)
   ={x_2^2 x_3^2 x_4^2\over (x_1-x_2)^2 (x_2-x_3)^2 (x_2-x_4)^2}
\label{secFmap}
\eea
where the $so(4,2)$ acts on the tensor product via the above twisted homomorphism, with the choice $ \lambda = x_2^2$. 
We will express this as 
\bea 
\cF_2 : V_+' \oplus V_+^{(p^2)} \otimes V_- \otimes V_- \rightarrow \mC 
\eea
The first factor  is written as $V_+'$ because the twist $ \alpha_{\lambda}$ is being applied. 
In the appendix \ref{appsec:QEOM-2}, we show that (\ref{secFmap}) indeed follows from 
the equivariance with respect to the twisted co-product. 
As in the discussion of the $x_1$ QEOM above, 
consider lifts of this map to 
\bea
\tilde \cF_2 :  V_+' \oplus \widetilde V_+ \otimes V_- \otimes V_- \rightarrow \mC 
\eea
In this case converting $P\cdot P$ into a differential operator is quite subtle.  
This is because there is $ x_2$ dependence in the vertex operator being  inserted 
at the second slot, but also $x_2$ dependence in the twist which determines the 
 map $ \widetilde  \cF_2$. We will leave the problem of resolving this subtlety as an 
 imporant techincal exercise for the future.

A similar discussion applies to $I^{(3)} $. 
There is an  $so(4,2)$ equivariant map 
\bea 
\cF_3 : V_+ \otimes V_+ \otimes \widetilde V_-^{(p^2) }  \otimes V_-'  \rightarrow \mC 
\eea
which gives the RHS of the quantum equation of motion for $x_3$. 
In this case we use a coproduct twisted on the last factor by the automorphism $\alpha_{x_3^{\prime 2}}$.
\bea 
 \cF_3(e^{-i P\cdot x_1}v^+\otimes e^{-iP\cdot x_2}v^+\otimes K_{ \mu} K_{ \mu} e^{iK\cdot x_3'}v^-\otimes e^{-iK\cdot x_4'}v^-) ={x_1^{\prime 2} x_2^{\prime 2} x_3^{\prime 2} \over (x_1'-x_3')^2 (x_2'-x_3')^2 (x_3'-x_4')^2}\cr
\eea 
%

\section{ Conclusions and  Future Directions } 
\label{sec:concl}

Much of our discussion of the four-point integral in four dimensions should 
generalize to the case of the three-point integral in six dimensions and the six-point integral in three dimensions, 
when we use the appropriate coordinate space propagators. 

\subsection{ Towards higher loops }  
\begin{figure}[ht]%
\begin{center}
\includegraphics[width=0.5\columnwidth]{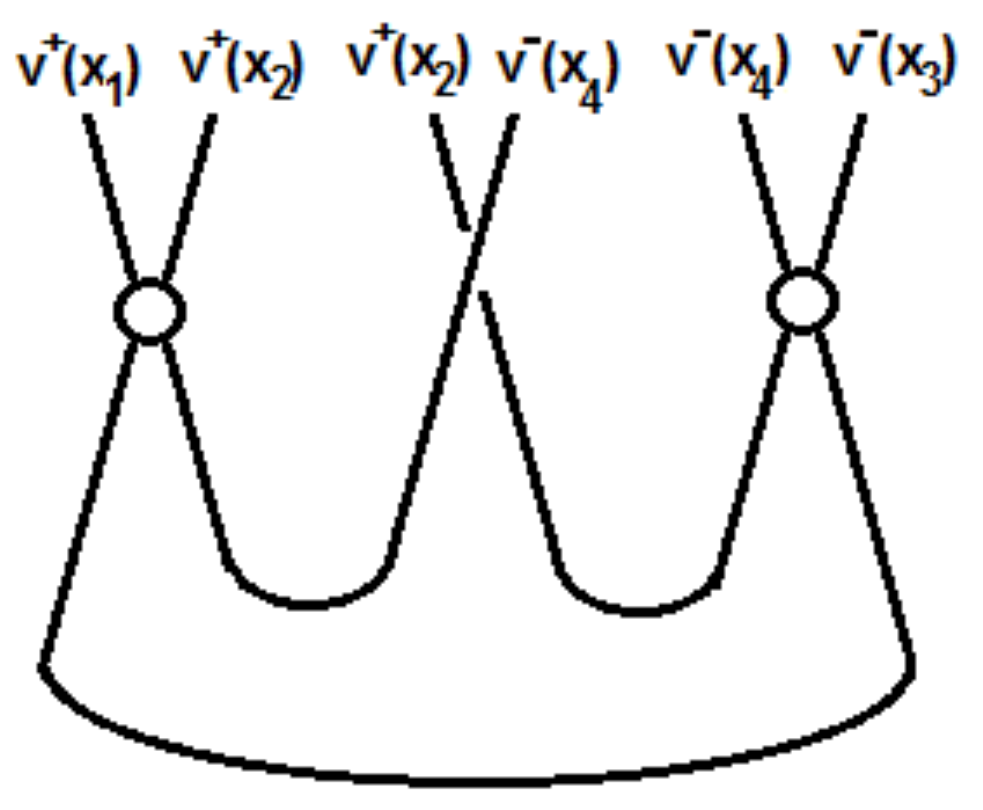}%
\caption{2-loop conformal integral}%
\label{fig:composingproj}%
\end{center}
\end{figure}

\begin{figure}[ht]%
\begin{center}
\includegraphics[width=0.9\columnwidth]{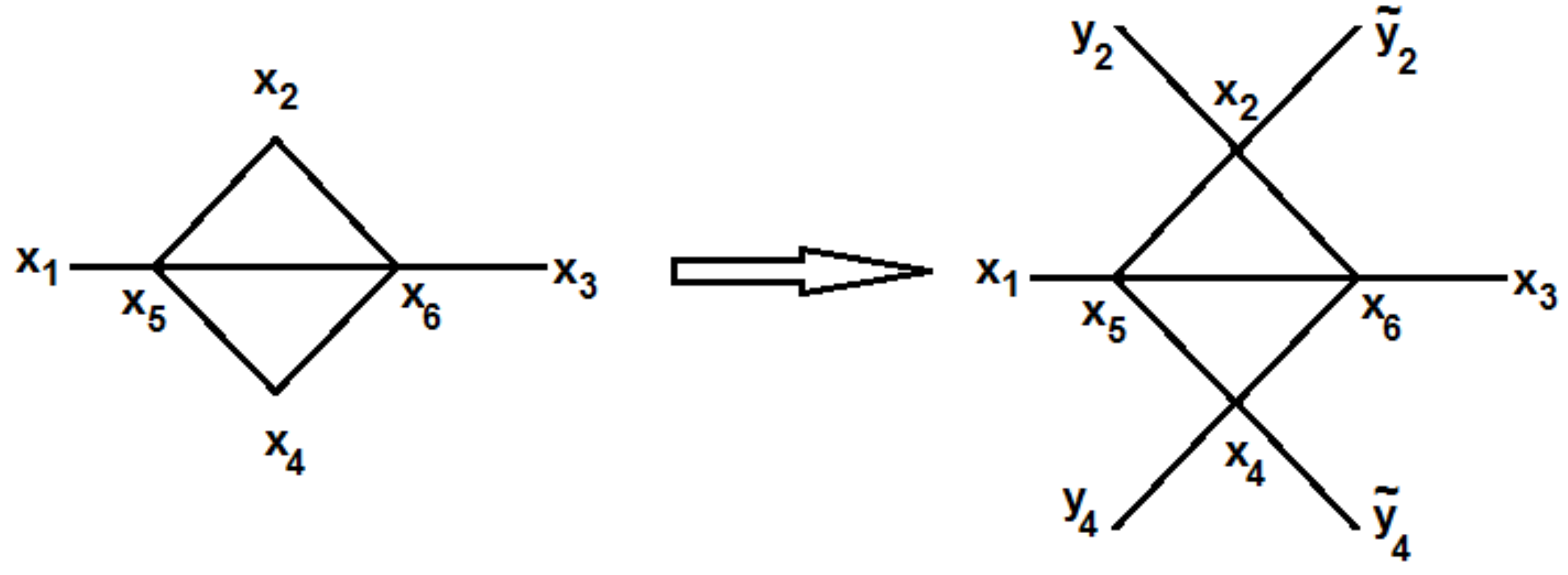}%
\caption{Closely-related-integral}%
\label{fig:extlegs}%
\end{center}
\end{figure}

We have focused attention on the case of the 1-loop conformal integral.
We outline how some key aspects of the discussion generalizes to 2-loops. 
The 2-loop integral is 
\bea\label{2-loop-integ}  
I_2 = \int d^4 x_5 d^4 x_6  { x_{56}^{-2} \over x_{15}^2 x_{25}^{2} x_{45}^2 x_{26}^2 x_{46}^2 x_{36}^2}  
\eea
where $x_{ij}^2 = ( x_i - x_j )^2 $. 
The exact answer is known. It has a term $ ( \log (s) )^2 F_2 ( s , t ) $. The term $F_2 ( s , t ) $ 
can be recovered from the HPEM of integration. Consider the order $ |x_1| < |x_2| < | x_3| < |x_4| $. 
This term arises from the integration range $ |x_2| < |x_5| < | x_6 |< |x_3|$. 
The expansion of the function $ F_2 ( s , t ) $ has an interpretation in terms of 
$so(4,2)$ equivariant maps indicated by  the diagram in Figure \ref{fig:composingproj}. 
It is an equivariant map acting on $e^{-iP\cdot x_1}v^+\otimes e^{-i P\cdot x_2}v^+\otimes e^{-i P\cdot x_2}v^+\otimes 
e^{iK\cdot x_3'}v^-\otimes e^{iK\cdot x_3'} v^-  \otimes e^{ i K\cdot x_4'} v^- $. The map is constructed by composing two projectors, one for each integration variable. There is an invariant map pairing 
two of  their indices, corresponding to the internal line. The projectors are the same ones we encountered in the 1-loop 
discussion $ V_+ \otimes V_+ \rightarrow V_{++} \rightarrow V_+ \otimes V_+$. 
It is also possible to modify the diagram, by attaching two external legs to each of $x_2, x_4 $ respectively (see Figure \ref{fig:extlegs}). 
In that case, the $x_2, x_4$ become integrated internal vertices. The resulting integral has a  fourth power of 
log which can be recovered from the HPEM method. The coefficient of this logarithmic term is interpreted 
in terms of a composition of four of the projectors, one for each internal vertex, and is closely related to the coefficient of the log-squared in $I_2$. 
We leave a more careful exposition of the  2-loop and higher loop cases to a forthcoming paper, but the above statements 
should be fairly plausible to the attentive  reader based on the discussion in this paper so far.

\subsection{TFT2 and renormalization} 

The original motivation for this work was to extend free CFT4/TFT2 to interacting theories.  
The discussion in this paper, developing the relation between Feynman integrals 
and equivariant maps, gives some useful clues in  this direction for the case of perturbative 
interacting CFTs. Concrete cases to consider are $ N=4$ SYM and the Wilson-Fischer fixed point. 
The connection between quantum equations of motion and indecomposable representations we have described 
should play a role. In the free CFT4/TFT2, we worked with a state space $ \bigoplus Sym^n ( V_+ \oplus V_- )$, 
where $V_+$ is the irreducible representation obtained from $\widetilde V_+ $ by quotienting out the $ V_+^{(p^2)} $. 
For the interacting CFT4, the state space should involve tensor products involving  $ \widetilde V_+ $ and $ \widetilde V_- $. 
There will be a coupling dependent quotient given by the quantum equations of motion. 
Once the correlators are computed at a given order in the perturbation expansion, we know that there is a 
renormalized formulation where these correlators are reproduced from local operators having 
dimensions shifted away from their values  in the free limit.  A TFT2 formulation of perturbative CFT
will presumably incorporate this renormalization in a sequence of TFT2s, one for each order of perturbation theory, 
such that the correlators computed at any stage of the sequence agree with each other. This will embed the 
renormalization for CFT4s in a TFT2 set-up: the benefit would be to keep, as much as possible, 
the conformal equivariance properties manifest in the process. 

\subsection{ Conformal blocks and CFT4/TFT2} 

The existence of a  TFT2 approach to CFT4 is made plausible by several known facts about CFT4.
CFT4  (and in fact more generally  CFT in any dimension)  has a distinctively algebraic flavour.
By the operator-state correspondence, local  operators correspond to definite representations
of $so(4,2)$. The spectrum of dimensions in the CFT4 along with the structure constants of 
the OPE determine the conformal field theory. The description of conformal blocks, which exploits Casimirs in a
central way, has a distinct similarity to projectors in representation theory (see for example \cite{DO0006,DO0011,SD1204} and more recently in a superconformal 
setting \cite{doohes}).  While these facts  strongly suggest  
the existence of a TFT2 formulation, the latter is not a trivial consequence. For example, to understand, from a 
purely representation theoretic point of view (as required in a TFT2 which by definition is about equivariant maps),  the 
fact that the OPE of $ \phi^2 $ with $ \phi^2$ in the free theory contains $ \phi^2$ \cite{CFT4TFT2} 
exploits, in an important way, the representation in TFT2 of the quantum field as a linear combination involving both $ V_+ $ and $ V_-$. The ordinary tensor product of $ V_{ ++}  \otimes V_{ ++} $ does not 
contain $ V_{++}$.   An interesting project in the CFT4/TFT2 programme is 
to understand in terms of equivariant maps, 
examples  in perturbative CFT4 of 4-point functions where factorization involves 
analogous OPEs, with both the positive dimension representations and their  negative dimension duals playing a role.

In special cases of OPEs of the type $\phi^2 \otimes \phi^2 \rightarrow \phi^4$, 
where the total number of fields in the intermediate operator is the same as in 
the external operators,  which 
have a direct analog in the tensor products of positive representations
without requiring negative energy representations in a crucial way, 
there should be a close relation between discussions of conformal blocks 
in the physics literature  \cite{DO0006,DO0011,SD1204} and the equivariant map interpretation 
of integrals developed in \cite{FL07} . 

\subsection{HPEM and the interaction/intertwiner connection} 

The HPEM has been an important  tool in our discussion of conformal integrals. 
It has enabled us to make the connection between $so(4,2)$ equivariant maps 
involving indecomposable representations and the quantum equations of motion. This connection 
 links a subtle aspect of representation theory with a consequence of the collision 
of interaction point with external vertex, a deep and generic property of interactions 
in quantum field theory. Using the HPEM the full integral was decomposed as a sum of terms $I^{(a)}$, each involving the collision of 
the interaction point with one of the external legs, and each associated with one of the QEOM. 
It will be very interesting to develop this physical picture for more general Feynman  integrals (not necessarily conformal), uncovering the interplay between the collision of interaction points, quantum equations of motion, and equivariance. 

As an example of a simple non-conformal integral consider in four dimensions 
an integration in coordinate space of an $n$-point scalar interaction (with $n\ne 4 $). 
To interpet in terms of  $so(4,2)$ equivariance, we would need to combine 
the scaling in spacetime, with a scaling of the coupling constant. In other words 
the equivariance would involve a ``twisted''  $so(4,2)'$ which combines  
the space-time $so(4,2)$ with an $so(2)$ scaling the coupling constant. 
Twistings which combine space-time symmetries with other symmetries are 
known to be useful in topological field theories.  The idea of employing a scaling 
of the coupling constant to arrive at a  generalized conformal symmetry 
was developed in \cite{JKY9810}.

\vspace*{1cm}

\begin{center} 
{ \bf Acknowledgements}
\end{center} 
 SR is supported by STFC Grant ST/J000469/1, String Theory, Gauge Theory, and Duality.
RdMK is supported by the South African Research Chairs Initiative of the Department of Science and Technology and the National Research Foundation. We thank Gabriele Travaglini for very useful  discussions at various stages of the project. We also thank Andi Brandhuber, Paul Heslop and Donovan Young for useful discussions.  SR thanks the Simons summer workshop for hospitality at the Simons  Centre for Geometry and Physics, and the Corfu workshop 
on non-commutative field theory and gravity for hospitality while part of this work was done. 

\begin{appendix} 
\section{ Basic formulae for HPEM (harmonic polynomial expansion method) } \label{basicf}

This Appendix summarizes the formulae used in the harmonic expansion method for integrals.   
This is also called the Gegenbauer Polynomial expansion technique. 

\noindent
{\it $so(4)$  harmonics : Notation } \\ 
We will expand the propagators using $so(4)$ spherical harmonics $Y^l_m(x) $. 
$x$ is a 4-vector $x^i$ in Euclidean space. 
The positive integer $l$ specifies a symmetric traceless $so(4)$ tensor with rank $l$. 
We will work with normalization
\bea 
{1\over 2\pi^2} \int dS^3 Y^{l_1}_{m_1}(x)Y^{l_2 }_{m_2}(x)=g_{m_1 m_2} ~ {\delta^{l_1, l_2}\over (l_1+1)} 
\eea
where $dS^3$ is the standard metric on the unit sphere.
We could work with more general bases where the $g$ factor depends on $l_1$, but we won't use this freedom. 
One convenient basis is  a real orthogonal basis for which
\bea 
&& g_{m_1 , m_2 }   = \delta_{ m_1 , m_2 } \cr  
&& Y_l^{m_1}  (x) = g^{m_1 m_2 } Y^l_{m_2} (x) = Y^l_{m_1} ( x )  \cr 
&& Y_{l}^{m} ( x ) = Y^l_m (x) 
\eea
Another convenient basis uses  the isomorphism $so(4) = su(2)\times su(2)$.  
A rank $l$ symmetric tensor specifies a representation of spins $({l\over 2},{l \over 2})$. 
The $so(4)$ state label $m$ is equivalent to a pair of $su(2)$ state labels $(m^L,m^R)$ each ranging from $-{l\over 2}$ 
to $ {l\over 2}$. 
If we work with a basis which diagonalize the $ J_3^L , J_3^R $, then the state label is equivalent to a pair of $su(2)$ 
labels $(m^L,m^R)$
\bea 
&& g_{m_1 , m_2 } =  \delta_{ m_1 , - m_2 } \cr 
&& Y_l^m (x) =  Y^{l}_{ - m}  (x) 
\eea
Using the explicit generators $J^L_i , J^R_i$ given in \cite{CFT4TFT2}, the charges $ ( J_3^L , J_3^R )$ for the basic variables are 
\bea
&&z = x_1+ix_2 \leftrightarrow ({1\over 2},{1\over 2})\cr
&&\bar z = x_1-ix_2 \leftrightarrow (-{1\over 2},-{1\over 2})\cr
&&w = x_3+ix_4 \leftrightarrow ({1\over 2},-{1\over 2})\cr
&&\bar w = x_3-ix_4 \leftrightarrow (-{1\over 2},{1\over 2})\label{nicebasis}
\eea
We find that 
\bea 
Y^{ l }_{ ( l/2 , l/2 ) } = ( x_1 + i x_2 )^{l}=z^l 
\eea
has normalization 
\bea 
{ 1 \over 2 \pi^2  } \int dS^3  ( Y^{ l }_{ (  {l \over 2}  ,{ l \over 2 } ) } )^*~~ 
Y^{ l }_{ ( { l \over 2 } ,{  l \over 2 }  ) } = { 1 \over ( l+1) } 
\eea
The remaining spherical harmonics are easily generated using the $su(2)$ lowering operators.

\noindent 
{\it Expansion of the exponential vertex operator } \\
The expansion of the exponential in terms of spherical harmonics is
\bea 
e^{ - i P\cdot x } v^+ = \sum_{ l ,  m } { (-i)^l  \over 2^l  l! } Y^l_m ( x  ) Y_l^{ m} ( P ) v^+ 
\eea
The invariant pairing  $\eta :V_+ \otimes V_-  \rightarrow \mC $ described in terms of harmonics in $P,K$ is
\bea 
\eta \left (Y^{l_1}_{m_1}(P)v^+,Y_{l_2}^{m_2}(K)v^-\right)=\delta_{l_1,l_2}\delta_{m_1,m_2}2^{2 l_1}(l_1!)^2 
\eea
The invariant pairing in terms of spherical harmonics in $x$ i.e. $\tilde\eta :H_+\otimes H_- \rightarrow \mC$ is
\bea 
\tilde \eta \left ( Y^{l_1}_{m_1}  ( x ) ,   { Y_{l_2}^{m_2}  ( x' ) \over x'^2 } \right ) = \delta_{ l_1 , l_2 } \delta_{ m_1 , m_2 }   
\eea
This can be written in terms of integration on the unit 3-sphere \cite{FL07}
\bea 
\tilde \eta \left ( Y^{l_1}_{m_1}  ( x ) ,   { Y_{l_2}^{m_2}  ( x' ) \over x'^2 } \right ) =
{ 1 \over 2 \pi^2} \int_{ S^3}  { Y_{l_2}^{m_2}  ( x' ) \over x'^2 } 
( x\cdot\partial_x +1 ) Y^{ l_1}_{m_1} ( x ) ) 
\eea
There are  $so(4,2)$ equivariant maps 
\bea 
&& f_+ : V_+ \rightarrow H_- \cr  
&& f_- : V_- \rightarrow H_+ 
\eea
which can be read off from the  expansion of the exponential 
\bea 
&& f_+ : Y^{l}_m ( P ) v^+ \rightarrow    { 2^l l! } (i)^l {  Y^l_m ( x' ) \over x'^2 } \cr 
&& f_- : Y^{l}_{m} (  K ) v^- \rightarrow  { 2^l l! } (-i)^l  Y^l_m ( x )
\eea
Combining the two maps 
$ f_+ \otimes f_- = f $ gives a map $ f : V_+  \otimes V_- \rightarrow  H_- \otimes H_+ $. We have $ \eta = f \circ \tilde \eta $.

\noindent 
{ \it Expansion of the 2-point function} \\ 
\bea
   {1\over |\vec{x}-\vec{y}|^2} && =\sum_{l=0}^\infty \sum_{ m  }
{1 \over y^{2}}Y^{ l }_{ m }(x )Y_{ l }^{ m }( y' )  ~~~ \hbox{ if} ~~~  |y| > |x| \cr 
&&  = \sum_{l=0}^\infty \sum_{ m  }
{1 \over x^{2}}Y^{ l }_{ m }(x' )Y_{ l }^{ m }( y )  ~~~ \hbox{ if} ~~~ |x| > |y| 
\label{sphIdent}
\eea
Using the addition theorem for spherical harmonics, the RHS  can be written in terms of Gegenbauer polynomials of $x\cdot y$. 
This is a well-known way of doing complicated integrals \cite{Kotikov}. 
We will not be writing expansions in terms of Gegenbauer polynomials, since our main purpose is to keep all four 
$ so(4,2)$'s associated with the external legs manifest, rather than finding an efficient way to do the integrals. 

\noindent 
{ \it Action of the quadratic $so(4,2)$ Casimir} \\ 
\bea
C_2=-{1\over 2}L_{MN}L^{MN}=D^2-4 D +\vec{P}\cdot\vec{K}+{1\over 2}M_{pq}M_{qp}
\eea
We will use the differential operator representation of the generators to compute the value of $C_2$ when acting 
on a product of a function of $r$ and a spherical harmonic.
As usual, we use vertex operators to obtain the differential opetator corresponding to a particular generator.
For example,
\bea
\vec{P}\cdot\vec{K}e^{-iP\cdot x}v^+ &=&
P_p\left(
2ix_p x\cdot {\partial\over\partial x}-i x^2 {\partial\over\partial x_p}+2i x_p\right)e^{-iP\cdot x}v^+ \cr
&=&
\left(
2ix_p x\cdot {\partial\over\partial x}-i x^2 {\partial\over\partial x_p}+2i x_p\right)
 i{\partial\over\partial x_p}\,\, e^{-iP\cdot x}v^+ 
\eea
To obtain the first equality for example, commute $K_\mu$ past the vertex operator $e^{-iP\cdot x}$ using the $so(4,2)$ algebra
and then use the fact that $K_\mu v^+=0$. Finally, express the result as a differential operator acting on the vertex 
operator. Similarly, we find
\bea
   M_{pq}=x_p{\partial\over\partial x_q}-x_q{\partial\over\partial x_p}
\eea
so that when acting on a power of $r$ times a spherical harmonic we have
\bea
    {1\over 2} M_{pq}M_{qp} (r^A Y^l_m (x)) = {r^A\over 2} M_{pq}M_{qp}  Y^l_m (x) = l (l+2)\, r^A Y^l_m (x)
\eea
We also have ($D=x\cdot {\partial\over\partial x}+1$ is shifted by 1 to account for the dimension of $v^+$)
\bea
   (D^2-4 D)\, r^A Y^ (x)=[(A+l-2)(A+l)-3]\, r^A Y^l_m (x)
\eea
Finally, consider $\vec{P}\cdot\vec{K}$
\bea
\vec{P}\cdot\vec{K}r^A Y^l_m (x)
=(A(A+2)+2Al-2 (A+l)^2 )\, r^A Y^l_m (x)
\eea
Thus we have
\bea
  C_2\, r^A Y^l_m (x)=-3\, r^A Y^l_m (x)
\eea
A very similar argument shows that
\bea
  C_2\,{\rm log} (r) Y^l_m (x)=-3 \,{\rm log} (r) Y^l_m (x)
\eea

\section{ Expansion of projector  using the exact answer } 
\label{App:expExact}

This section extends the discussion of section \ref{CIandFL} by providing some of the details behind the expansion.
We want to study the limit $s\rightarrow 0,$  $t\rightarrow 1$, i.e. $ u \rightarrow 0 $ where $t=1+u$. 
In this limit $ \lambda \rightarrow 0 $ and $ \rho \rightarrow \infty $. 
The coefficient of the $\log s$ is
\bea 
F_1 = {1\over\lambda}\log\left( (1+\rho s) \over ( 1 + \rho^{-1} t^{-1} ) \right ) 
\eea
Using Mathematica, we expand in $s$ and simplify the function of $u$ appearing at each order of $s$, to obtain
\bea 
&& F_1 = { \log ( 1 + u ) \over u } + { s\over u^3 } \left ( 2 u - ( 2 +u ) \log ( 1 +u ) \right ) 
+ { s^2 \over u^5 } \left ( 3u ( 2 + u ) - 6 + 6u + u^2 ) \log ( 1 +u ) \right )   \cr 
&& + { s^3 \over 3  u^{ 7} }  \left (    u ( 60 + 60 u + 11 u^2 ) - 3 ( 20  + 30 u + 12 u^2 + u^3 ) \log ( 1 + u ) \right )
\eea
After subsequently expanding in powers of $u$, we have 
\bea 
&& F_1 = \left ( -1 + { u \over 2 } - { u^2 \over 3 } + { u^3 \over 4 }  - \cdots \right ) 
      + { s} \left ( { -1 \over 6  } + { u \over 6 } - { 3 u^2 \over 20 } + \cdots \right ) + 
     s^2 \left ( { -1 \over 30  } + { u \over 20 } - {  2u^2 \over 35 } + \cdots \right ) \cr 
&& 
 + s^3 \left (    { -1 \over 140 } + { u \over 70 } - { 5 u^2 \over 252 }    + \cdots       \right ) 
+ \cdots 
\eea
The term at  $ s^k $ is 
\bea\label{formexp}  
{s^k \over u^{2k +1}}\left( uP_{k}(u)-Q_k(u)\log (1+u)\right)  
\eea
where $P_k (u)$ is a polynomial of degree $k-1$ and $Q_k (u)$ is a polynomial of degree $k$, both with positive coefficients. 
The polynomials have the property that the expansion at $s^{ k} $ is regular at $ u =0$.
In other words 
\bea\label{PolyPQ} 
 u P_{k}  ( u ) - Q_k ( u ) \log ( 1 + u ) 
\eea
only contains powers $u^n$ with $n>2k$. 
This gives $2k+1$ equations constraining the $ (2k+1)$ unknown coefficients in $ P_k $ and $Q_k$. 
These equations do not determine the overall normalization of the two polynomials. 
This is determined by observing that the leading coefficient in $P_k (u)$ is
\bea 
 {(2k)!\over (k!)^2} 
\eea
which is $(k +1)$ times the $k$'th Catalan number. Writing
\bea 
Q_k ( u ) = \sum_{ l =0 }^{ k}  b_l u^l  
\eea
we find the linear system of equations 
\bea\label{LinSym}  
\sum_{ l=0}^{ k } b_l { (-1)^{  l } \over ( i -l ) } = 0
\eea
as $i$ ranges from $ k +1 $ to $2k $. 
These come from the requirement that the coefficient of $x^i$ vanishes  in (\ref{PolyPQ}). 
These equations allow us to solve $b_1, b_2,\cdots ,b_k$ in terms of $b_0$. 
For example, when $k=5$ we have 
\bea 
b_1={5b_0\over 2}, \quad b_2={20b_0\over 9},\quad  b_3={5b_0\over 6},\quad  b_4 ={5b_0\over 42}, \quad 
b_5={b_0 \over 252} 
\eea
Interestingly  the Catalan number $252$ comes from solving this system of linear equations.  
To solve (\ref{LinSym}) for any $k$, define $I=i-k$, where the range of $I$ is $1\le I\le k$ and we have
\bea 
\sum_{ l =0 }^{ k } M_{ I l } b_l  = 0\label{nicer}
\eea
where 
\bea 
M_{ I l } = { ( -1)^l \over ( I + k - l ) } 
\eea
The index $l$ ranges over $k+1$ values. We can rewrite (\ref{nicer}) as
\bea 
M_{ I 0 } b_0 = - \sum_{ l =1 }^k  M_{ I l } b_l 
\eea
Define $\hat M_{ab} = M_{ab}$ with $a,b=1,...,k$. 
Using Mathematica to study a few examples, we have verified that $\hat M$ is invertible, so that we can write
\bea 
b_l = \sum_{ I } \hat M^{-1}_{ J I  } M_{ I 0 } b_0 
\eea
For a specific choice of $k$, it is easy to generate the inverses of $\hat M$ in Mathematica and hence to generate the $b_I / b_0$. 
For $k=2 ... 7 $, we find 
\bea 
&& b_I  / b_0 = \{ 1 , 1/6 \} \cr 
&& b_I / b_0 = \{ 3/2,3/5,1/20 \} \cr
&& b_I / b_0 = \{  2, 9/7, 2/7, 1/70 \} \cr 
&& b_I/b_0 = \left\{\frac{5}{2},\frac{20}{9},\frac{5}{6},\frac{5}{42},\frac{1}{252}\right\}\cr
&& b_I/b_0 = \left\{3,\frac{75}{22},\frac{20}{11},\frac{5}{11},\frac{1}{22},\frac{1}{924}\right\} \cr
&& b_I/b_0 = \left\{\frac{7}{2},\frac{63}{13},\frac{175}{52},\frac{175}{143},\frac{63}{286},\frac{7}{429},\frac{1}{3432}\right\}\cr 
&& 
\eea
Using these numerical results from Mathematica and the Online Encyclopaedia of Integer Sequences\cite{OEIS}, we find
\bea 
&& b_1 ( k )/ b_0  = k/2 \cr 
&& b_k  ( k ) / b_0 = { k!^2 \over ( 2k) ! } \cr 
&& b_{k-1} ( k ) /b_0  = {   k!^2  \over (2k)! }  k ( k+1) \cr 
&& b_{ k-2} ( k ) /b_0  = {   k!^2  \over (2k)! }{  ( k-1 )  k ( k +1 ) ( k+2 ) \over 4 } \cr 
&& b_{ k-3 }   ( k ) /b_0  = {   k!^2  \over (2k)! }{ k \choose 3 } { k+3 \choose 3 } \cr 
&& b_{ k - m   }  ( k ) / b_0  = {   k!^2  \over (2k)! } { k \choose m  } { k+ m \choose m  } 
\eea
This gives the general formula for the $b$-coefficients. 
We also know $b_0 = {  ( 2k) ! \over  k!^2 } $, so that we have
\bea 
&& b_1 ( k ) = k/2 {  ( 2k) ! \over  k!^2 }  \cr 
&& b_k  ( k )  = 1  \cr 
&& b_{k-1} ( k )   =  k ( k+1) \cr 
&& b_{ k-2} ( k )  = {  ( k-1 )  k ( k +1 ) ( k+2 ) \over 4 } \cr 
&& b_{ k-3 }   ( k )  = { k \choose 3 } { k+3 \choose 3 } \cr 
&& b_{ k - m   }  ( k )   = { k \choose m  } { k+ m \choose m  } 
\eea 
Lets us now consider the polynomial $P_k (u)$
\bea
   P_k(u)=\sum_{i=1}^k a_i u^{i-1}
\eea
By looking at powers $u^i$ in (\ref{PolyPQ}) for $1 \le i \le k $, we obtain the linear equations
\bea 
a_i  -  \sum_{ l=0}^{ i-1} b_l { (-1)^{ i - l } \over ( i -l ) } = 0 
\eea
This gives the $a_i$ as sums of binomial coefficients, using the formula for $b$ above. 
We can again numerically work out the $a$-coefficients for small values of $k$ and then read off the
analytic formulas from the patterns we find. 
For the first few values of $k$ we find
\bea 
&& k=2 :\qquad \vec a = \{ a_1 , ... a_k \} = \{ 6 , 3 \} \cr  
&& k=3 :\qquad \vec a = { 1 \over 3 }  \times  \{ 60 , 60 , 11 \} \cr 
&& k=4 :\qquad \vec a  = { 5 \over 6 }  \times   \{ 84 , 126 , 52 , 5 \} \cr 
&& k=5 :\qquad \vec a =  { 1 \over 30 } \times \{ 7560 , 15120 , 9870 , 2310 , 137 \} \cr 
&& k=6 :\qquad \vec a = { 7 \over 10 } \times \{ 13720 , 3300 , 2960 , 1140 , 174 , 7 \} 
\eea
which leads to 
\bea 
&& a_1 (k) = { ( 2 k)! \over (k!)^2 } \cr 
&& a_k ( k ) = 2 h(k) = 2 \sum_{ l=1}^{ k } { 1 \over l } 
\eea
$h(k)$ is the k-th harmonic number. 
Thus, the $a_i (k)$ interpolate between Catalan numbers and harmonic numbers. 
In this way, the Catalan numbers, along with the form (\ref{formexp}), has determined 
all the coefficients in the double Taylor expansion around $s,u=0$. 

\label{SumClebschs} 
\subsection{A summation formula for products of $su(2)$ Clebschs from Feynman Integrals} 

Our discussion implies a summation formula for products of 
$su(2)$-Clebsh Gordan coefficients, since these coefficients appear in the multiplication of 
spherical harmonics which enter the definition of the $so(4,2)$  equivariant  map $\cM$.
Indeed, equating the result for the coefficient of the log obtained from HPEM to the coefficient of the same
log appearing in the exact result, we have
\begin{equation} 
\boxed{ 
\begin{aligned}
~~~~& \sum_{l_i=0}^\infty \sum_{m_i}\sum_{t=0}^\infty\sum_{p}
\delta_{l_1+l_2,l_3+l_4} Y^{l_1}_{ - m_1}(x_1) Y^{l_2}_{-m_2}(x_2) Y^{l_3 }_{ m_3}(x'_3) Y^{l_4}_{m_4}(x'_4) 
\times {C^{{l_1},{l_2 };{t }}_{ m_1, m_2;  p   } C^{l_3 ,l_4 ;  t   }_{ -m_3, - m_4;  - p } \over t+1} ~~~~ \\
&\qquad ={ x_3^2 x_4^2 \over x_{13}^2 x_{24}^2 }
{1\over\lambda}\log\left((1+\rho s)\over (1+\rho^{-1}t^{-1})\right) ~~~~~
\label{smrl}
\end{aligned} 
} \\ 
\end{equation} 
Noting that the structure constants of the multiplication of $so(4)$-covariant  harmonics on $S^3$ 
can be written in terms of Clebsch-Gordan coefficients of $ su(2)$
\begin{equation*} 
\mathlarger{ \mathlarger{C^{ ~l_1 , ~ l_2 ~;~~~ l}_{ m_1 , m_2~ ; ~m_1 + m_2 }} }  = 
\left \langle \begin{tabular}{cc} 
${ l_1 \over 2 }$ &   ${ l_2 \over 2 }$ \\   
${ m_1^L  \over  2 }$ & ${ m_2^L \over 2  }$  
                  \end{tabular} \Bigg \vert      
 \begin{tabular}{c} ${ l \over 2 }$ \\
                            ${ m_1^L + m_2^L \over 2 }$ \end{tabular}  \right \rangle 
\left \langle \begin{tabular}{cc} 
${ l_1 \over 2 }$ &   ${ l_2 \over 2 }$ \\   
${ m_1^R  \over  2 }$ & ${ m_2^R \over 2  }$  
                  \end{tabular} \Bigg \vert      
 \begin{tabular}{c} ${ l \over 2 }$ \\
                            ${ m_1^R + m_2^R \over 2 }$ \end{tabular}  \right \rangle 
\end{equation*} 
we see that (\ref{smrl}) is a highly nontrivial sum rule for $su(2)$-Clebsh Gordan coefficients.

To check this sum rule, it is useful to use the basis which diagonalizes the $(J_3^L,J_3^R)$ sub-algebra of $su(2)\times su(2)$,
described in (\ref{nicebasis}).
Using this basis we can easily determine the coefficients of monomials of a specific form appearing on both side of (\ref{smrl}).
The simplest monomial arises from the terms in the sum with $ t=0$ and 
\bea 
&& Y^{{n_1\over 2}}_{{n_1\over 2},{n_1 \over 2}}(x_1)Y^{{n_2\over 2}}_{{n_2\over 2},{n_2\over 2}}(x_2)
     Y^{{n_3 \over 2}}_{{-n_3\over 2},{-n_3\over 2}}(x_1)Y^{{n_4\over 2}}_{{-n_4\over 2},{-n_4\over 2}}(x_4)  
= z_1^{n_1} z_2^{n_2 } \bar z_3^{n_3 } \bar z_4^{n_4 } 
\eea
In this extremal case, the Clebsch Gordan coefficients are 1, so that the monomial 
$z_1^{n_1}z_2^{n_2}\bar z_3^{n_3} \bar z_4^{n_4}$ has coefficient ${1\over n_1+n_2+1}$.
To recover this coefficient from the RHS of (\ref{smrl}), note that
\bea 
( x_i- x_j)^2 &=& 1 - 2 x_i\cdot x_j + x_i^2 x_j^{2}\cr 
&=&1-z_i \bar z_j - \bar z_i z_j - w_i \bar w_j - \bar w_i w_j + ( z_i \bar z_i + w_i \bar w_i ) ( z_j \bar z_j + w_j \bar w_j)  
\eea
Inserting this into the RHS of (\ref{smrl}) and expanding as described at the start of this Appendix, we can obtain the coefficient of 
any given monomial.
In particular, we verify that $z_1^{n_1}z_2^{n_2}\bar z_3^{n_3} \bar z_4^{n_4}$ has coefficient ${1\over n_1+n_2+1}$.
Next consider
\bea
   Y^{n_1\over 2}_{{n_1\over 2},{n_1-2\over 2}}(x_1)
   Y^{n_2\over 2}_{{n_2-2\over 2},{n_2\over 2}}(x_2)
   Y^{n_3\over 2}_{-{n_3\over 2},-{n_3-2\over 2}}(x_3)
   Y^{n_4\over 2}_{-{n_4-2\over 2},-{n_4\over 2}}(x_4) \cr\cr
= \sqrt{n_1 n_2 n_3 n_4}z_1^{n_1-1}\bar{w}_1z_2^{n_2-1}w_2\bar{z}_3^{n_3-1}w_3\bar{z}_4^{n_4-1}\bar w_4
\eea
which invloves terms in the sum with $t=0$ and $t=1$.
We need two Clebsch Gordan coefficients
\bea
C^{{n_1\over 2},{n_2\over 2},{n_1+n_2\over 2}}_{\left({n_1\over 2},{n_1-2\over 2}\right),\left({n_2-2\over 2},{n_2\over 2}\right),\left({n_1+n_2-2\over 2},{n_1+n_2-2\over 2}\right)}
={\sqrt{n_1 n_2}\over n_1+n_2}
\eea
\bea
C^{{n_1\over 2},{n_2\over 2},{n_1+n_2-2\over 2}}_{\left({n_1\over 2},{n_1-2\over 2}\right),\left({n_2-2\over 2},{n_2\over 2}\right),\left({n_1+n_2-2\over 2},{n_1+n_2-2\over 2}\right)}
={\sqrt{n_1 n_2}\over n_1+n_2}
\eea
which easily follow from the following $su(2)$ coefficients
\bea
C^{{n_1\over 2},{n_2\over 2},{n_1+n_2\over 2}}_{{n_1\over 2},{n_2-2\over 2},{n_1+n_2-2\over 2}}
=\sqrt{ n_2\over n_1+n_2}
\eea
\bea
C^{{n_1\over 2},{n_2\over 2},{n_1+n_2-2\over 2}}_{{n_1\over 2},{n_2-2\over 2},{n_1+n_2-2\over 2}}
=-\sqrt{ n_1\over n_1+n_2}
\eea
Explicit formulae for the $su(2)$ Clebsch Gordan coefficients are available in \cite{Wiki-Table-Clebsch}.
Thus, the coefficient of $z_1^{n_1-1}\bar{w}_1z_2^{n_2-1}w_2\bar{z}_3^{n_3-1}w_3\bar{z}_4^{n_4-1}\bar w_4$ is
\bea
\sqrt{n_1 n_2 n_3 n_4}\left[
C^{{n_1\over 2},{n_2\over 2},{n_1+n_2\over 2}}_{\left({n_1\over 2},{n_1-2\over 2}\right),\left({n_2-2\over 2},{n_2\over 2}\right),\left({n_1+n_2-2\over 2},{n_1+n_2-2\over 2}\right)}{1\over n_1+n_2+1}\right.\cr
\left.+C^{{n_1\over 2},{n_2\over 2},{n_1+n_2-2\over 2}}_{\left({n_1\over 2},{n_1-2\over 2}\right),\left({n_2-2\over 2},{n_2\over 2}\right),\left({n_1+n_2-2\over 2},{n_1+n_2-2\over 2}\right)}{1\over n_1+n_2}
\right]\cr
={2 n_1 n_2 n_3 n_4 \over (n_1+n_2-1) (n_1+n_2) (n_1+n_2+1)}
\eea
This again agrees with the coefficient obtained by expanding the RHS of (\ref{smrl}).

\label{appsec:QEOM}
\section{Equivariant maps related to Quantum Equations of Motion}

This section gives the explicit evaluation of the maps $\cF_1$ and $\cF_2$ which are introduced in section \ref{sec:QEOM-indec}.

\label{appsec:QEOM-1}
\subsection{Quantum equation of motion for $x_1$} 
When we evaluate the map $\cF_1$ with the exponential states inserted, we get an expression which 
has a well defined expansion at small $ x_1 , x_2' , x_3' , x_4'$.
Set $h(x_1,x_2',x_3',x_4')={\mathcal{F}}_1 (e^{-iP\cdot x_1}v^{+++}\otimes e^{iK\cdot x_2'}v^{-}\otimes e^{iK\cdot x_3'}v^-\otimes e^{iK\cdot x_4'}v^-)$.
Then
\bea
{\partial h\over\partial x^\mu_1}
&=&-i \cF_1(P_\mu e^{-iP\cdot x_1}v^{+++}\otimes e^{iK\cdot x_2'}v^{-}
\otimes e^{iK\cdot x_3'}v^-\otimes e^{iK\cdot x_4'}v^-)\cr
&=&i{\mathcal{F}}_1 (e^{-iP\cdot x_1}v^{+++}\otimes P_\mu e^{iK\cdot x_2'}v^{-}
\otimes e^{iK\cdot x_3'}v^-\otimes e^{iK\cdot x_4'}v^-)\cr
&&+i{\mathcal{F}}_1 (e^{-iP\cdot x_1}v^{+++}\otimes e^{iK\cdot x_2'}v^{-}
\otimes P_\mu e^{iK\cdot x_3'}v^-\otimes e^{iK\cdot x_4'}v^-)\cr
&&+i{\mathcal{F}}_1 (e^{-iP\cdot x_1}v^{+++}\otimes e^{iK\cdot x_2'}v^{-}
\otimes e^{iK\cdot x_3'}v^-\otimes P_\mu e^{iK\cdot x_4'}v^-)
\label{frstattmpt}
\eea
where the last line follows from the $so(4,2)$ invariance of $\cF_1$.
Now,
\bea
   P_\mu e^{iK\cdot x_3'}v^-
&=&\left(2i x_{3\mu}'x_3'\cdot {\partial\over\partial x_3'}-ix_3^{\prime 2}{\partial\over\partial x_{3\mu}'}+2ix_{3\mu}'\right)
e^{iK\cdot x_3'}v^-\cr
&=&i x_3^2 \,\, {\partial\over\partial x_3^\mu} \,\, {1\over x_3^2} e^{iK\cdot x_3'}v^-
\eea
so that (\ref{frstattmpt}) becomes
\bea
{\partial h\over\partial x^\mu_1}
&=&- x_2^2 \,\, {\partial\over\partial x_2^\mu} \,\, ({1\over x_2^2}h)
- x_3^2 \,\, {\partial\over\partial x_3^\mu} \,\, ({1\over x_3^2}h)
- x_4^2 \,\, {\partial\over\partial x_4^\mu} \,\, ({1\over x_4^2}h)
\eea
which can be written as
\bea
x_2^2 x_3^2 x_4^2 \left({\partial\over\partial x_1^\mu}+{\partial\over\partial x_2^\mu}+{\partial\over\partial x_3^\mu}
+{\partial\over\partial x_4^\mu}\right) \left[ {h\over x_2^2 x_3^2 x_4^2}\right]=0
\eea
This proves the map is only a function of the differences $x_i^\mu -x_j^\mu$.
Next, consider
\bea
\left( x_{1\mu}{\partial\over\partial x_{1\nu}}-x_{1\nu}{\partial\over\partial x_{1\mu}}\right)h
&=&\cF_1(M_{\mu\nu}e^{-iP\cdot x_1}v^{+++}\otimes e^{iK\cdot x_2'}v^{-}
\otimes e^{iK\cdot x_3'}v^-\otimes  e^{iK\cdot x_4'}v^-)\cr
&=&-{\mathcal{F}}_1 (e^{-iP\cdot x_1}v^{+++}\otimes M_{\mu\nu}e^{iK\cdot x_2'}v^{-}
\otimes e^{iK\cdot x_3'}v^-\otimes e^{iK\cdot x_4'}v^-)\cr
&-&{\mathcal{F}}_1 (e^{-iP\cdot x_1}v^{+++}\otimes e^{iK\cdot x_2'}v^{-}
\otimes M_{\mu\nu} e^{iK\cdot x_3'}v^-\otimes  e^{iK\cdot x_4'}v^-)\cr
&-&{\mathcal{F}}_1 (e^{-iP\cdot x_1}v^{+++}\otimes e^{iK\cdot x_2'}v^{-}
\otimes e^{iK\cdot x_3'}v^-\otimes M_{\mu\nu} e^{iK\cdot x_4'}v^-)
\eea
which after a little algebra can be written as
\bea
x_2^2 x_3^2 x_4^2 \left(x_{1\mu}{\partial\over\partial x_{1\nu}}-x_{1\nu}{\partial\over\partial x_{1\mu}}
+x_{2\mu}{\partial\over\partial x_{2\nu}}-x_{2\nu}{\partial\over\partial x_{2\mu}}
+x_{3\mu}{\partial\over\partial x_{3\nu}}-x_{3\nu}{\partial\over\partial x_{3\mu}}\right.\cr
\left. +x_{4\mu}{\partial\over\partial x_{4\nu}}-x_{4\nu}{\partial\over\partial x_{4\mu}}
\right) \left[ {h\over x_2^2 x_3^2 x_4^2}\right]=0
\eea
This proves the map is only a function of the magnitudes of the differences $|x_i^\mu -x_j^\mu|$.
Next, consider
\bea
\left(x_1\cdot {\partial\over\partial x_1}+3\right)h
&=&\mathcal{F}_1 (e^{-iP\cdot x_1}v^{+++}\otimes e^{iK\cdot x_2'}v^{-}\otimes e^{iK\cdot x_3'}v^-\otimes  e^{iK\cdot x_4'}v^-)\cr
&=&-\mathcal{F}_1 (e^{-iP\cdot x_1}v^{+++}\otimes D e^{iK\cdot x_2'}v^{-}\otimes e^{iK\cdot x_3'}v^-\otimes  e^{iK\cdot x_4'}v^-)\cr
&-&\mathcal{F}_1(e^{-iP\cdot x_1}v^{+++}\otimes e^{iK\cdot x_2'}v^{-}\otimes D e^{iK\cdot x_3'}v^-\otimes  e^{iK\cdot x_4'}v^-)\cr
&-&\mathcal{F}_1 (e^{-iP\cdot x_1}v^{+++}\otimes e^{iK\cdot x_2'}v^{-}\otimes e^{iK\cdot x_3'}v^-\otimes  D e^{iK\cdot x_4'}v^-)
\eea
This can be rewritten as
\bea
x_2^2 x_3^2 x_4^2 \left(x_1\cdot {\partial\over\partial x_1}+
x_2\cdot {\partial\over\partial x_2}+x_3\cdot {\partial\over\partial x_3}+x_4\cdot {\partial\over\partial x_4}+12\right)
\left[ {h\over x_2^2 x_3^2 x_4^2}\right]=0
\eea
This tells us the degree of the dependence on $|x_i^\mu -x_j^\mu|$.
Thus, at this stage we know that
\bea
 {h\over x_2^2 x_3^2 x_4^2}=
{A\over |x_1-x_2|^{2\alpha}|x_1-x_3|^{2\beta}|x_1-x_4|^{2\gamma}|x_2-x_3|^{2\delta}|x_2-x_4|^{2\eta}|x_3-x_4|^{2\tau}}
\eea
and $\alpha+\beta+\gamma+\delta+\eta+\tau = 6$.
Thus, the map has now been reduced to determining 7 numbers.
To determine these numbers start by noting that at $x_1= 0$ and $x_2'=x_3'=x_4'= 0$ we have
\bea
h(0,0,0,0)= \cF_1(v^{+++}\otimes v^{-}\otimes v^-\otimes  v^-)=1
\eea
i.e. $h$ has no singularities and takes the constant value 1.
This implies that $\delta=\eta=\tau=0$, $\alpha=\beta=\gamma=1$ and $A=1$, which proves
(\ref{fmapexpected}).

\subsection{ Quantum equation of motion for $x_2$ } \label{appsec:QEOM-2}
Using invariance of the map and the $so(4,2)$ algebra,we can easily prove (\ref{secFmap}).
Towards this end, recall some results which follow from the invariant pairing
\bea
\eta (e^{-iP\cdot x_1}v^+,e^{iK\cdot x_2'}v^-)
=\sum_{n,m=0}^\infty {i^m (-1)^n\over n!m!}x_1^{\alpha_1}\cdots x_1^{\alpha_n}
x_2^{\prime\beta_1}\cdots x_2^{\prime\beta_m}T_{\alpha_1\cdots\alpha_n\beta_1\cdots\beta_m}
\eea
where
\bea
T_{\alpha_1\cdots\alpha_n\beta_1\cdots\beta_m}&=&
\eta(P_{\alpha_1}\cdots P_{\alpha_n}v^+,K_{\beta_1}\cdots K_{\beta_m}v^-)\cr
&=& (-1)^n \eta(v^+,P_{\alpha_n}\cdots P_{\alpha_1}K_{\beta_1}\cdots K_{\beta_m}v^-)\cr
&=&(-1)^m\eta(K_{\beta_m}\cdots K_{\beta_1}P_{\alpha_1}\cdots P_{\alpha_n}v^+,v^-)
\eea
and the last two lines above follow from the $so(4,2)$ invariance of the pairing.
Now, setting
\bea
K_{\beta_m}\cdots K_{\beta_1}P_{\alpha_1}\cdots P_{\alpha_n}v^+ 
=\delta_{nm}t_{\alpha_1\cdots\alpha_n\beta_1\cdots\beta_m} v^+\label{frsres}
\eea
\bea
P_{\alpha_n}\cdots P_{\alpha_1}K_{\beta_1}\cdots K_{\beta_m}v^-
=\delta_{nm}t_{\alpha_1\cdots\alpha_n\beta_1\cdots\beta_m} v^-\label{secres}
\eea
we find
\bea
T_{\alpha_1\cdots\alpha_n\beta_1\cdots\beta_m}=t_{\alpha_1\cdots\alpha_n\beta_1\cdots\beta_m} (-1)^m
\eea
and
\bea
\sum_{n=0}^\infty { (-1)^n\over n!m!}x_1^{\alpha_1}\cdots x_1^{\alpha_n}
x_2^{\prime\beta_1}\cdots x_2^{\prime\beta_n}t_{\alpha_1\cdots\alpha_n\beta_1\cdots\beta_n}
={1\over 1-2x_2'\cdot x_1+x_1^2 x_2^{\prime 2}}\label{thrres}
\eea
We will make use of (\ref{frsres}), (\ref{secres}) and (\ref{thrres}) below.
Consider the complete expansion
\bea
&&\cF_2(e^{iP\cdot x_1}v^+\otimes e^{-iP\cdot x_2}v^{+++}\otimes e^{iK\cdot x_3'}v^-\otimes e^{iK\cdot x_4'}v^-)\cr
&=&
\sum_{n_1,n_3,n_4=0}^\infty {1\over n_1! n_3! n_4!}
\cF_2 ((iP\cdot x_1)^{n_1}v^+\otimes e^{-iP\cdot x_2}v^{+++}\otimes (iK\cdot x_3')^{n_3}v^-\otimes (iK\cdot x_4')^{n_4}v^-)
\cr
&&\eea
Expand the remaining exponential and use equivariance of the map to transfer the $P\cdot x_2$ factors into the
other three slots. 
Due to the twisting, when we move $P\cdot x_2$ into the first slot we get
\bea
   \alpha (P\cdot x_2)={1\over x_2^2}K\cdot x_2 = K\cdot x_2'
\eea  
For a given term in the sum (i.e. a given $n_1,n_3,n_3$) only a specific power of $P\cdot x_2$ from the expansion of the
exponential in slot 2 will contribute.
Keeping only this power we have
\bea
&&\cF_2(e^{iP\cdot x_1}v^+\otimes e^{-iP\cdot x_2}v^{+++}\otimes e^{iK\cdot x_3'}v^-\otimes e^{iK\cdot x_4'}v^-)\cr
&=&
\sum_{n_1,n_3,n_4=0}^\infty {1\over n_1! n_3! n_4! (n_1+n_3+n_4)!}\cr
&&\times
\cF_2((iP\cdot x_1)^{n_1}v^+\otimes (-iP\cdot x_2)^{n_1+n_3+n_4}v^{+++}
\otimes (iK\cdot x_3')^{n_3}v^-\otimes (iK\cdot x_4')^{n_4}v^-)\cr
\cr
&=&
\sum_{n_1,n_3,n_4=0}^\infty {1\over n_1! n_3! n_4! (n_1+n_3+n_4)!} {(n_1+n_3+n_4)!\over n_1! n_3! n_4!}
(-1)^{n_1+n_3+n_4}\cr
&&\times
\cF_2((\alpha (P\cdot x_2))^{n_1}(P\cdot x_1)^{n_1}v^+\otimes v^{+++}
\otimes (P\cdot x_2)^{n_3}(K\cdot x_3')^{n_3}v^-\otimes (P\cdot x_2)^{n_4}(K\cdot x_4')^{n_4}v^-)\cr
\cr
&=&
\sum_{n_1}^\infty {(-1)^{n_1}\over (n_1!)^2}
x_1^{\alpha_1}\cdots x_1^{\alpha_{n_1}}x_2^{\prime\beta_1}\cdots x_2^{\prime\beta_{n_1}}
t_{\alpha_1\cdots \alpha_{n_1}\beta_1\cdots \beta_{n_1}}\cr
&&\times
\, \sum_{n_3}^\infty {(-1)^{n_3}\over (n_1!)^2}
x_2^{\gamma_1}\cdots x_2^{\gamma_{n_3}}x_3^{\prime \delta_1}\cdots x_3^{\prime \delta_{n_3}}
t_{\gamma_1\cdots\gamma_{n_3}\delta_1\cdots\delta_{n_3}}\cr
&&\times\sum_{n_4}^\infty {(-1)^{n_4}\over (n_1!)^2}
x_2^{\mu_1}\cdots x_2^{\mu_{n_4}}x_4^{\prime \nu_1}\cdots x_4^{\prime \nu_{n_4}}
t_{\mu_1\cdots\mu_{n_4}\nu_1\cdots\nu_{n_4}}
\cF_2(v^+\otimes v^{+++}\otimes v^- \otimes  v^-)\cr\cr
&=&{1\over 1-2x_1\cdot x_2'+x_1^2x_2^{\prime 2}}
{1\over 1-2x_2\cdot x_3'+x_2^2x_3^{\prime 2}}
{1\over 1-2x_2\cdot x_4'+x_2^2x_4^{\prime 2}}\cr\cr
&=&{x_2^2 x_3^3 x_4^2 \over (x_1-x_2)^2(x_2-x_3)^2(x_2-x_4)^2}
\eea
which proves the result (\ref{secFmap}).

\end{appendix}


\begin{thebibliography}{}  


\bibitem{Feyncount}
  R.~de Mello Koch and S.~Ramgoolam,
  ``Strings from Feynman Graph counting : without large N,''
  Phys.\ Rev.\ D {\bf 85} (2012) 026007
  [arXiv:1110.4858 [hep-th]].

\bibitem{DMW}
  R.~de Mello Koch, S.~Ramgoolam and C.~Wen,
  ``On the refined counting of graphs on surfaces,''
  Nucl.\ Phys.\ B {\bf 870} (2013) 530
  [arXiv:1209.0334 [hep-th]].


\bibitem{QuivCalc}
  J.~Pasukonis and S.~Ramgoolam,
  ``Quivers as Calculators: Counting, Correlators and Riemann Surfaces,''
  JHEP {\bf 1304} (2013) 094
  [arXiv:1301.1980 [hep-th]].

\bibitem{KimFrob}
  Y.~Kimura,
  ``Multi-matrix models and Noncommutative Frobenius algebras obtained from symmetric groups and Brauer algebras,''
  Commun.\ Math.\ Phys.\  {\bf 337} (2015) 1,  1
  [arXiv:1403.6572 [hep-th]].

\bibitem{JRR}
  V.~Jejjala, S.~Ramgoolam and D.~Rodriguez-Gomez,
  ``Toric CFTs, Permutation Triples and Belyi Pairs,''
  JHEP {\bf 1103} (2011) 065
  [arXiv:1012.2351 [hep-th]].

\bibitem{Tom-complex}
  T.~W.~Brown,
  ``Complex matrix model duality,''
  Phys.\ Rev.\ D {\bf 83} (2011) 085002
  [arXiv:1009.0674 [hep-th]].

\bibitem{CJR}
  S.~Corley, A.~Jevicki and S.~Ramgoolam,
  ``Exact correlators of giant gravitons from dual N=4 SYM theory,''
  Adv.\ Theor.\ Math.\ Phys.\  {\bf 5} (2002) 809
  [hep-th/0111222].

\bibitem{GiGraOsc}
  R.~d.~M.~Koch, M.~Dessein, D.~Giataganas and C.~Mathwin,
  ``Giant Graviton Oscillators,''
  JHEP {\bf 1110} (2011) 009
  [arXiv:1108.2761 [hep-th]].

\bibitem{DoubCos}
  R.~de Mello Koch and S.~Ramgoolam,
  ``A double coset ansatz for integrability in AdS/CFT,''
  JHEP {\bf 1206} (2012) 083
  [arXiv:1204.2153 [hep-th]].

\bibitem{Geloun:2013kta} 
  J.~Ben Geloun and S.~Ramgoolam,
  ``Counting Tensor Model Observables and Branched Covers of the 2-Sphere,''
  arXiv:1307.6490 [hep-th].

\bibitem{minzar} 
  J.~A.~Minahan and K.~Zarembo,
  ``The Bethe ansatz for N=4 superYang-Mills,''
  JHEP {\bf 0303} (2003) 013
  [hep-th/0212208].

\bibitem{bks0303}
  N.~Beisert, C.~Kristjansen and M.~Staudacher,
  ``The Dilatation operator of conformal N=4 superYang-Mills theory,''
  Nucl.\ Phys.\ B {\bf 664} (2003) 131
  [hep-th/0303060].

\bibitem{CFT4TFT2} 
  R.~de Mello Koch and S.~Ramgoolam, ``CFT$_4$ as $so(4,2)$-invariant TFT$_2$,''
  Nucl.\ Phys.\ B {\bf 890} (2014) 302
  [arXiv:1403.6646 [hep-th]].

\bibitem{Atiyah}
M.~Atiyah, ``Topological quantum field theory,'' Publ. Math. I.H.E.S, tome 68 (1988), 175-186.

\bibitem{mooseg} 
  G.~W.~Moore and G.~Segal,
  ``D-branes and K-theory in 2D topological field theory,''
  hep-th/0609042.

\bibitem{malda}
  J.~M.~Maldacena,
  ``The Large N limit of superconformal field theories and supergravity,''
  Int.\ J.\ Theor.\ Phys.\  {\bf 38} (1999) 1113
   [Adv.\ Theor.\ Math.\ Phys.\  {\bf 2} (1998) 231]
  doi:10.1023/A:1026654312961
  [hep-th/9711200].

\bibitem{gkp}
  S.~S.~Gubser, I.~R.~Klebanov and A.~M.~Polyakov,
  Phys.\ Lett.\ B {\bf 428} (1998) 105
  doi:10.1016/S0370-2693(98)00377-3
  [hep-th/9802109].

\bibitem{witten} 
  E.~Witten,
  Adv.\ Theor.\ Math.\ Phys.\  {\bf 2} (1998) 253
  [hep-th/9802150].


\bibitem{KKN1410}
  Y.~Kazama, S.~Komatsu and T.~Nishimura,
  ``Novel construction and the monodromy relation for three-point functions at weak coupling,''
  JHEP {\bf 1501} (2015) 095
   [JHEP {\bf 1508} (2015) 145]
  [arXiv:1410.8533 [hep-th]].

\bibitem{JKPS1410}
  Y.~Jiang, I.~Kostov, A.~Petrovskii and D.~Serban,
  ``String Bits and the Spin Vertex,''
  Nucl.\ Phys.\ B {\bf 897} (2015) 374
  [arXiv:1410.8860 [hep-th]].

\bibitem{BKV1505}
  B.~Basso, S.~Komatsu and P.~Vieira,
  ``Structure Constants and Integrable Bootstrap in Planar N=4 SYM Theory,''
  arXiv:1505.06745 [hep-th].

\bibitem{ADGN0510}
  L.~F.~Alday, J.~R.~David, E.~Gava and K.~S.~Narain,
  ``Towards a string bit formulation of N=4 super Yang-Mills,''
  JHEP {\bf 0604} (2006) 014
  [hep-th/0510264].

\bibitem{FLM1405}
  L.~Freidel, R.~G.~Leigh and D.~Minic,
  ``Quantum Gravity, Dynamical Phase Space and String Theory,''
  Int.\ J.\ Mod.\ Phys.\ D {\bf 23} (2014) 12,  1442006
  [arXiv:1405.3949 [hep-th]].
  
  \bibitem{FLM1502}
  L.~Freidel, R.~G.~Leigh and D.~Minic,
  ``Metastring Theory and Modular Space-time,''
  JHEP {\bf 1506} (2015) 006
  [arXiv:1502.08005 [hep-th]].
  
\bibitem{banks}
  T.~Banks,
  Phys.\ Rev.\ D {\bf 52} (1995) 2462
  doi:10.1103/PhysRevD.52.2462
  [hep-th/9503145].

\bibitem{moore1}
  A.~Losev, G.~W.~Moore, N.~Nekrasov and S.~Shatashvili,
  Nucl.\ Phys.\ B {\bf 484} (1997) 196
  doi:10.1016/S0550-3213(96)00612-8
  [hep-th/9606082].



\bibitem{moore2} 
  A.~Losev, G.~W.~Moore, N.~Nekrasov and S.~Shatashvili,
  Nucl.\ Phys.\ Proc.\ Suppl.\  {\bf 46} (1996) 130
  doi:10.1016/0920-5632(96)00015-1
  [hep-th/9509151].




\bibitem{Costa:2011mg} 
  M.~S.~Costa, J.~Penedones, D.~Poland and S.~Rychkov,
  ``Spinning Conformal Correlators,''
  JHEP {\bf 1111}, 071 (2011)
  [arXiv:1107.3554 [hep-th]].

\bibitem{Dixon-review-1310}
  L.~J.~Dixon,
  ``A brief introduction to modern amplitude methods,''
  arXiv:1310.5353 [hep-ph].

\bibitem{DHSS07}
  J.~M.~Drummond, J.~Henn, V.~A.~Smirnov and E.~Sokatchev,
  ``Magic identities for conformal four-point integrals,''
  JHEP {\bf 0701} (2007) 064
  [hep-th/0607160].

\bibitem{DKS0707}
  J.~M.~Drummond, G.~P.~Korchemsky and E.~Sokatchev,
  ``Conformal properties of four-gluon planar amplitudes and Wilson loops,''
  Nucl.\ Phys.\ B {\bf 795} (2008) 385
  [arXiv:0707.0243 [hep-th]].

\bibitem{FL07}
I. Frenkel and M. Libine, ``Quaternionic analysis, representation theory and physics,'' Advances in Mathematics 218, no. 6 (2008): 1806-1877, arXiv:0711.2699 [math.RT].

\bibitem{Aizawa:2014yqa} 
  N.~Aizawa and V.~K.~Dobrev,
  ``Intertwining Operator Realization of anti de Sitter Holography,''
  doi:10.1016/S0034-4877(15)30002-1
  arXiv:1406.2129 [hep-th].

\bibitem{DO0006}
  F.~A.~Dolan and H.~Osborn,
  ``Implications of N=1 superconformal symmetry for chiral fields,''
  Nucl.\ Phys.\ B {\bf 593} (2001) 599
  [hep-th/0006098].

\bibitem{DO0011}
  F.~A.~Dolan and H.~Osborn,
  ``Conformal four point functions and the operator product expansion,''
  Nucl.\ Phys.\ B {\bf 599} (2001) 459
  [hep-th/0011040].

\bibitem{SD1204} 
  D.~Simmons-Duffin,
  ``Projectors, Shadows, and Conformal Blocks,''
  JHEP {\bf 1404}, 146 (2014)
  [arXiv:1204.3894 [hep-th]].

\bibitem{UssDavPLB305}
  N.~I.~Usyukina and A.~I.~Davydychev,
  ``Exact results for three and four point ladder diagrams with an arbitrary number of rungs,''
  Phys.\ Lett.\ B {\bf 305} (1993) 136.

\bibitem{UssDav9307}
  N.~I.~Usyukina and A.~I.~Davydychev,
  ``Some exact results for two loop diagrams with three and four external lines,''
  Phys.\ Atom.\ Nucl.\  {\bf 56} (1993) 1553
   [Yad.\ Fiz.\  {\bf 56N11} (1993) 172]
  [hep-ph/9307327].

\bibitem{Rych-Tan}
  S.~Rychkov and Z.~M.~Tan,
  ``The $\epsilon$-expansion from conformal field theory,''
  J.\ Phys.\ A {\bf 48} (2015) 29,  29FT01
  [arXiv:1505.00963 [hep-th]].

\bibitem{DO0209} 
  F.~A.~Dolan and H.~Osborn,
  ``On short and semi-short representations for four-dimensional superconformal symmetry,''
  Annals Phys.\  {\bf 307} (2003) 41
  [hep-th/0209056].

\bibitem{KMMR0512} 
  J.~Kinney, J.~M.~Maldacena, S.~Minwalla and S.~Raju,
  ``An Index for 4 dimensional super conformal theories,''
  Commun.\ Math.\ Phys.\  {\bf 275} (2007) 209
  [hep-th/0510251].

\bibitem{bouffe}
  M.~Bianchi, P.~J.~Heslop and F.~Riccioni,
  ``More on La Grande Bouffe,''
  JHEP {\bf 0508} (2005) 088
  [hep-th/0504156].
  
\bibitem{Lib1309}
  M.~Libine,
  ``The Two-Loop Ladder Diagram and Representations of U(2,2),''
  arXiv:1309.5665 [math.RT].

\bibitem{Lib1407}
  M.~Libine,
  ``The Conformal Four-Point Integrals, Magic Identities and Representations of U(2,2),''
  arXiv:1407.2507 [math.RT].
  ,VJS11,DoFlo08
  
  \bibitem{Kotikov} 
A.~V.~Kotikov,
``The Gegenbauer polynomial technique: The Evaluation of a class of Feynman diagrams,''
Phys.\ Lett.\ B {\bf 375}, 240 (1996)
[hep-ph/9512270].

  
\bibitem{Gaberdiel0105}
 M.~R.~Gaberdiel,
  ``Fusion rules and logarithmic representations of a WZW model at fractional level,''
  Nucl.\ Phys.\ B {\bf 618} (2001) 407
  [hep-th/0105046].

\bibitem{VJS11}
  R.~Vasseur, J.~L.~Jacobsen and H.~Saleur,
  ``Indecomposability parameters in chiral Logarithmic Conformal Field Theory,''
  Nucl.\ Phys.\ B {\bf 851} (2011) 314
  [arXiv:1103.3134 [math-ph]].
%
 \bibitem{DoFlo08}
  A.~L.~Do and M.~Flohr,
  ``Towards the construction of Local Logarithmic Conformal Field Theories,''
  Nucl.\ Phys.\ B {\bf 802} (2008) 475
  [arXiv:0710.1783 [hep-th]].

\bibitem{DOcas}
F.~A.~Dolan and H.~Osborn,
``Conformal partial waves and the operator product expansion,''
Nucl.\ Phys.\ B {\bf 678}, 491 (2004)
[hep-th/0309180].

\bibitem{Dolan2005}
  F.~A.~Dolan,
  ``Character formulae and partition functions in higher dimensional conformal field theory,''
  J.\ Math.\ Phys.\  {\bf 47} (2006) 062303
  [hep-th/0508031].

\bibitem{Heiden1980}
  W.~Heidenreich,
  ``Tensor Products of Positive Energy Representations of SO(3,2) and SO(4,2),''
  J.\ Math.\ Phys.\  {\bf 22} (1981) 1566.

\bibitem{doohes}
  R.~Doobary and P.~Heslop,
  ``Superconformal partial waves in Grassmannian field theories,''
  arXiv:1508.03611 [hep-th].

\bibitem{JKY9810}
  A.~Jevicki, Y.~Kazama and T.~Yoneya,
  ``Generalized conformal symmetry in D-brane matrix models,''
  Phys.\ Rev.\ D {\bf 59} (1999) 066001
  [hep-th/9810146].

\bibitem{OEIS}
See ``The On-Line Encyclopedia of Integer Sequences,'' available at http://oeis.org/.

\bibitem{Wiki-Table-Clebsch} 
https://en.wikipedia.org/wiki/Table\_of\_Clebsch\%E2\%80\%93Gordan\_coefficients


\end{thebibliography}
\end{document}